\documentclass[prd,twocolumn,preprintnumbers,footinbib,tightenlines,nofootinbib,superscriptaddress]{revtex4}

\usepackage[english]{babel}
\usepackage{amsmath,amssymb,amsfonts, bm,bbm,slashed, subdepth}
\usepackage{graphicx}
\usepackage[sort&compress]{natbib}
\usepackage{xcolor}
\usepackage[normalem]{ulem}
\usepackage{hyperref}
\usepackage{cleveref}
\usepackage{enumerate}
\usepackage{setspace}
\usepackage{booktabs, tabularx}
\usepackage{units}
\usepackage{placeins}
\usepackage{multirow}
\usepackage{mathtools}

\usepackage{floatrow}

\newfloatcommand{capbtabbox}{table}[][\FBwidth]

\usepackage{blindtext}

\begin{document}

\preprint{DESY 19-137}
\preprint{IPMU19-0113}
\title{A Practical and Consistent Parametrization of Dark Matter Self-Interactions}

\author{Xiaoyong Chu}
\email{xiaoyong.chu@oeaw.ac.at }
\affiliation{Institute of High Energy Physics, Austrian Academy of Sciences, Nikolsdorfer Gasse 18, 1050 Vienna, Austria
}

\author{Camilo Garcia-Cely}
\email{camilo.garcia.cely@desy.de}
\affiliation{Deutsches Elektronen-Synchrotron DESY, Notkestrasse 85,
22607 Hamburg, Germany}
\author{Hitoshi Murayama}
\email{hitoshi@berkeley.edu,\,\,hitoshi.murayama@ipmu.jp, \,\, Hamamatsu Professor}
\affiliation{Department of Physics, University of California, Berkeley, CA 94720, USA}
\affiliation{Kavli Institute for the Physics and Mathematics of the
  Universe (WPI), University of Tokyo,
  Kashiwa 277-8583, Japan}
\affiliation{Ernest Orlando Lawrence Berkeley National Laboratory, Berkeley, CA 94720, USA}
\affiliation{Deutsches Elektronen-Synchrotron DESY, Notkestrasse 85,
22607 Hamburg, Germany}

\begin{abstract}

Self-interacting dark matter has been proposed to explain the apparent mass deficit in astrophysical small-scale  halos, while observations from galaxy clusters suggest that the corresponding
cross section depends on the velocity. 
Accounting for this is often believed to be highly model-dependent with studies mostly focusing on scenarios with light mediators.   
Based on the effective-range formalism, in this work we point out a model-independent approach which accurately approximates the velocity dependence of the self-interaction cross section with only two parameters, in addition to the dark matter mass. 
We illustrate how this parametrization can be simultaneously interpreted in various well-motivated scenarios, including self-interactions induced by 
Yukawa forces,  Breit-Wigner resonances and  bound states. We investigate the astrophysical implications and discuss how the approximation can be improved in certain special regimes where it works poorly.

 \end{abstract}

\maketitle

\tableofcontents

\section{Introduction}

In contrast to the ordinary substances found on Earth, more than three quarters of the matter in the Universe is not made of protons, neutrons, or electrons.  This is the so-called dark matter (DM) and identifying its particle nature is one of the chief goals of particle physics and cosmology today. 
According to the $\Lambda$CDM model~\cite{Aghanim:2018eyx}, which accurately describes the Universe at cosmological scales,  DM interacts very weakly with normal
matter and it was cold and collisionless during the formation of structures in the early universe. 
Although DM can be treated as collisionless particles at large scales, non-gravitational  DM scatterings can still occur in the dense central regions of small-scale halos such as those of dwarf or low-surface-brightness galaxies. This is  the self-interacting dark matter (SIDM) hypothesis, which was proposed~\cite{Spergel:1999mh} to explain the seeming discrepancies between observations of the smallest DM halos that we can currently observe and certain predictions of the $\Lambda$CDM model; see \cite{Tulin:2017ara, Bullock:2017xww} for recent reviews. 

The aforementioned  discrepancies  can be explained if DM elastically scatters with a cross section per unit of mass as large as several cm$^2/$g when it moves at approximately $10$\,km/s, {\it i.e.}\/, roughly the DM velocity dispersion  in small-scale objects~\cite{Dave:2000ar,Vogelsberger:2012ku,Rocha:2012jg,Peter:2012jh,Elbert:2014bma,Fry:2015rta}. Meanwhile, recent studies on halo dynamics at cluster scales provide upper bounds on the self-interaction cross section of around $\unit[0.2]-\unit[1]{cm^2/g}$~\cite{Randall:2007ph,Kaplinghat:2015aga, Robertson:2016xjh, Bondarenko:2017rfu, 2018ApJ...853..109E, Harvey:2018uwf}, which  are associated with typical DM  velocities of the order  of $\unit[10^3]{km/s}$. 

A natural question then arises: how can these observations be interpreted in terms of the properties of the DM particle?  One possibility is to postulate a specific DM model, and translate the previous velocity-dependent cross section in terms of masses and couplings. For instance, this has been done for scenarios where DM interacts by means of a light mediator~\cite{Feng:2009hw, Buckley:2009in,  Tulin:2012wi} and for models in which DM resonantly self-scatters~\cite{Chu:2018fzy, Ibe:2009mk, Duch:2017nbe}. Nevertheless, from the phenomenological point of view, this is not very practical, not only because the cross sections typically depend in a complicated way on the model parameters, but also because there  are a myriad of SIDM scenarios. The essence of this work is to propose a simple parametrization of the DM self-interaction cross section, which  approximates with great accuracy the velocity-dependent effects, and most importantly, which interpolates the predictions of different DM scenarios allowing to establish comparisons among them. More precisely, here we advocate the use of the effective-range theory as a model-independent way to study the velocity dependence of SIDM. Notice that this approach has been adopted in concrete models of DM before~\cite{MarchRussell:2008tu, Braaten:2013tza, Cline:2013zca, Braaten:2018xuw,Mahbubani:2019pij}.

The effective-range approach 
was  formulated~\cite{Bethe:1949yr, Blatt:1949zz} as an effort to explain  the non-relativistic scattering of neutrons by protons. 
Based on simple assumptions from quantum mechanics, this approach suggests that the scattering observables 
can be parametrized in terms of two quantities: the scattering length $a$, and the effective range $r_e$. 
While being very predictive, the effective-range theory does not demand a precise knowledge of the underlying interactions among the colliding particles, apart from the requirement that the scattering force must vanish at sufficiently large distances. In fact, due to this, it can  describe the non-relativistic scattering induced by contact interactions, light mediators, and  Breit-Wigner resonances, among others.

A brief historical remark may be helpful to readers.\footnote{Hans Bethe personally described the history in a YouTube video \url{https://www.youtube.com/watch?v=hbcQMG2XpTI.} } In late 1940s people proposed different models to describe nucleon-nucleon scattering cross sections at low energies.  After many explicit calculations, it became clear that only two parameters are relevant, independent of details of models.  Schwinger came up with a proof why that was the case in an unpublished lecture note.  Blatt and Jackson~\cite{Blatt:1949zz} showed with more calculations that indeed only two parameters were necessary to explain the data.  Then Bethe~\cite{Bethe:1949yr} came up with a simple and elegant proof (reproduced in Appendix~\ref{app:ERT}) to understand this observation.  The flip side of this remarkable simplicity is that we gain very little information on the detailed model from the data.  Bethe wrote ``{\it practically no information could be obtained, from classical scattering experiments, on the shape of the potential.}''  On the other hand, for the purpose of describing the impact of self-interaction among dark matter particles in various halos, this is a boon; we need to specify only two parameters (and the mass of dark matter) in order to simulate the impact of self-interactions without the need for dealing with explicit models.  This is why we propose the use of the effective range theory for the study of SIDM.

This paper is organized as follows. In Section~\ref{sec:ERT}, we review the effective-range approach, and explain how it can be useful to describe DM self-scattering. In Section~\ref{sec:SIDMv},  we discuss the implications of this approach for describing the velocity-dependence of DM self-interactions in astrophysical halos in a model-independent way. In Section~\ref{sec:models}, we investigate how to relate the  scattering length and  effective range   to parameters of concrete DM models.   
 In Section~\ref{sec:gen}, we  propose a concrete method on how the effective-range approach can be extended or improved in some cases where it fails.   Section~\ref{sec:conclusions}  provides the final conclusions and future prospects.

\begin{figure*}[t]
\includegraphics[trim=0.cm 0cm 0.0cm 0.cm,clip,height=0.227\textheight]{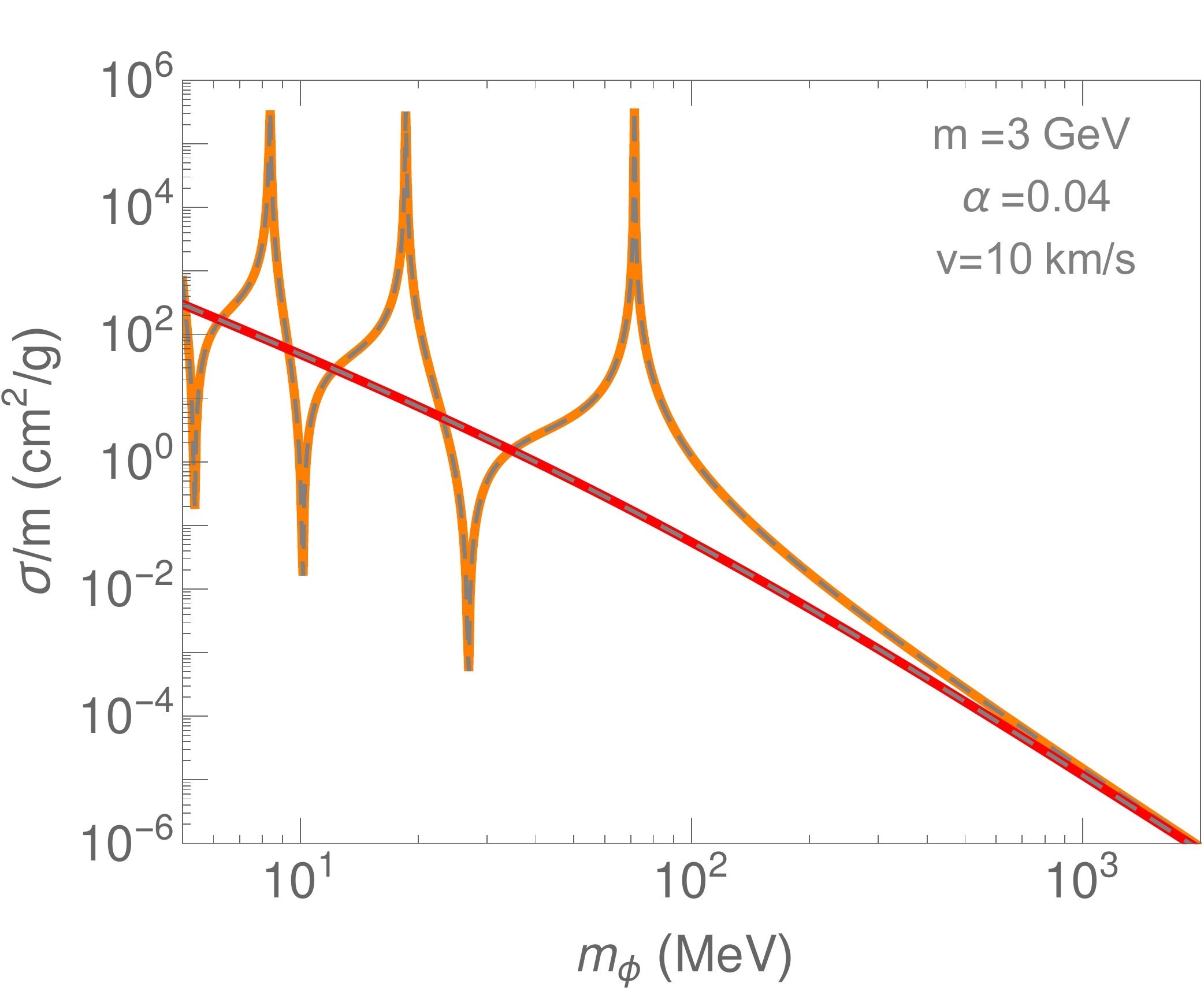}
\includegraphics[trim=1.3cm 0cm 0.0cm 0.cm,clip,height=0.225\textheight]{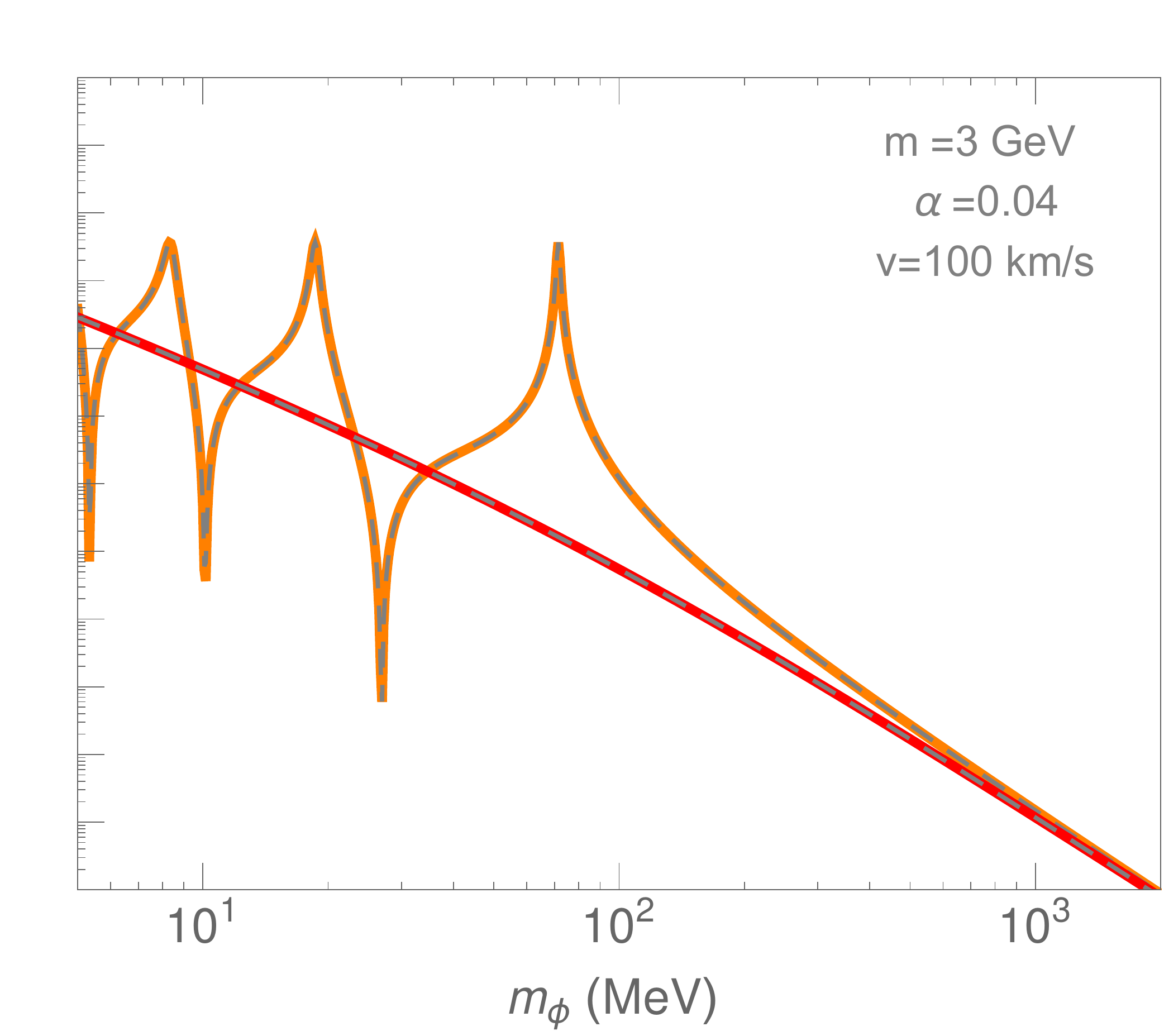}
\includegraphics[trim=1.3cm 0cm 0.0cm 0.cm,clip,height=0.225\textheight]{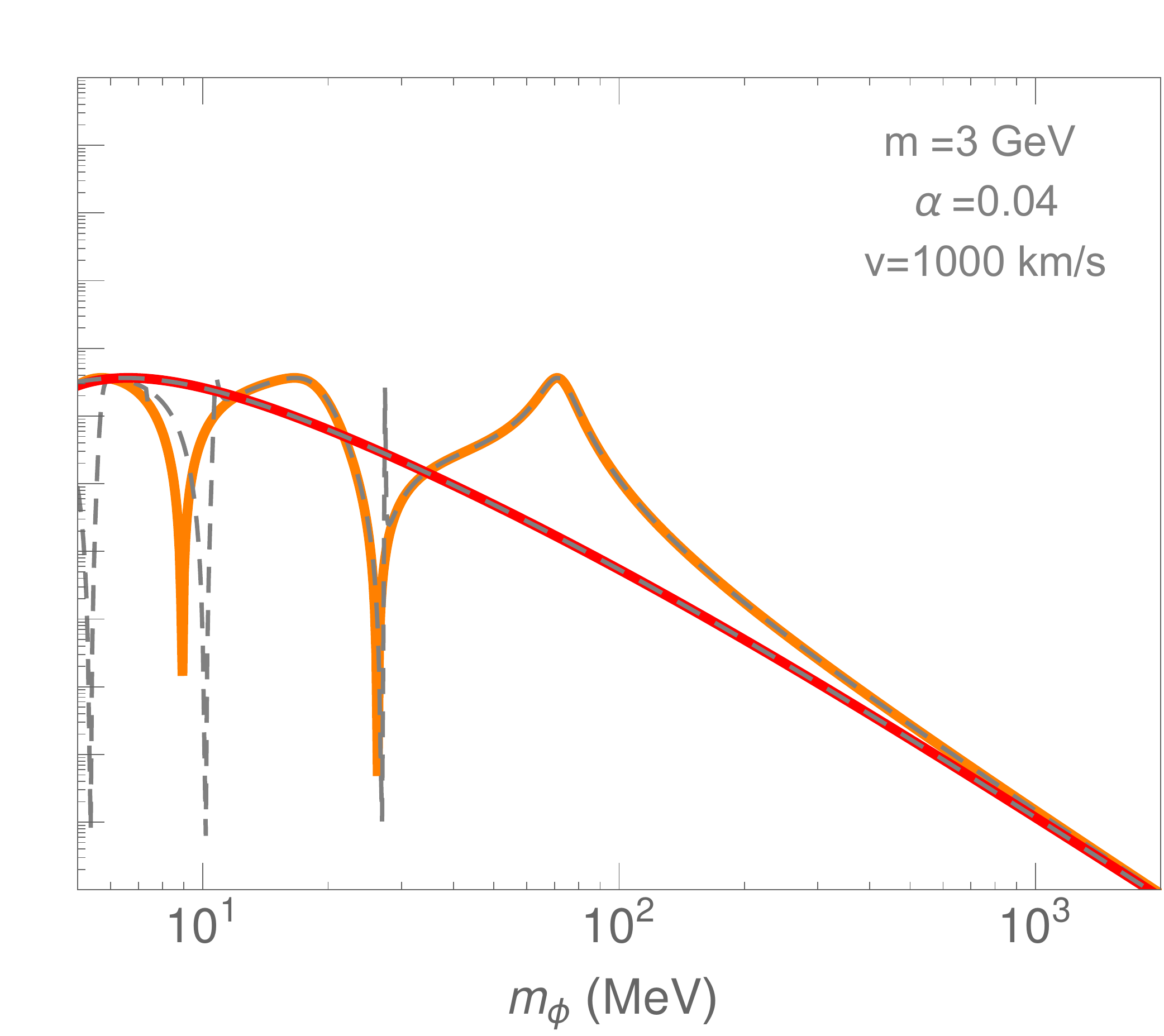}
\caption{
Comparison of the numerical S-wave cross section per unit mass (solid) against the effective-range approximation (dashed). Orange and red lines correspond to the numerical results for attractive and repulsive forces, respectively. 
Note that the mass here is the actual mass of the scattering particles, not the reduced mass.
}
\label{fig:alpha}
\end{figure*}

\section{The effective-range formalism}
\label{sec:ERT}

Before focusing on SIDM, we will first introduce the effective-range approximation. In any collision the differential cross section $d\sigma/d\Omega$ determines the scattering rate and describes the   velocity  dependence of the process. Up to a possible  symmetry factor, it  is given by $d\sigma =|f(k,\theta)|^2 d\Omega$,  where $f(k,\theta)$ is the scattering amplitude with $k$ being the incoming momentum.\footnote{Here and below, we separate the center-of-mass motion and hence $k$ is the relative momentum of two particles that scatter.  The orbital angular momentum $\ell$ below is defined in the center-of-mass frame.  Note also that we are dealing with low velocities, less than about $10^{-2} c$, which are typical in halos, so the use of non-relativistic quantum mechanics is justified.}
 For collisions with definite orbital angular momentum, $\ell$, the amplitude is proportional to the Legendre polynomial, $P_\ell(\cos\theta)$, with the corresponding  coefficient defining the partial-wave amplitude, $f_\ell(k)$.  More precisely
\begin{align}
 f(k,\theta)&=\sum^\infty_{\ell=0} (2l+1) f_\ell(k) P_\ell \left(\cos \theta\right)\,,\\
\text{~with~~}& f_\ell(k) \equiv  \frac{e^{2i \delta_\ell(k)}-1}{2 i k}=\frac{1}{k\,( \cot \delta_\ell(k)-i) }\,.
\label{eq:fkdef}
\end{align}
The second relation defines the phase shift,  $\delta_\ell(k)$ for the $\ell$ partial wave.
While the precise value of  $\delta_\ell(k)$ must be obtained by solving the Schr\"odinger equation describing the scattering process, this phase shift always satisfies some general requirements. 
For instance, it must be real if inelastic processes are absent. In this work, we are concerned with  elastic scatterings in astrophysical halos, as a result, unless stated otherwise, we will assume that inelastic processes are relatively weaker and take $\delta_\ell$ real (see Section~\ref{subsec:inelastic} for how to include the inelastic processes). 

Another requirement on the phase shift is that, for finite-range interactions, the function $k^{2\ell+1} \cot \delta_\ell (k)$ must be analytic at $k=0$ (for more details see Appendix~\ref{app:ERT}).  
The effective-range approximation~\cite{Bethe:1949yr, Blatt:1949zz}  consists in neglecting the high-order terms in the corresponding expansion in $k^2$,  so that 
\begin{equation}
k^{2\ell+1} \cot \delta_\ell (k) \simeq -\frac{1}{a_\ell^{2\ell+1}}+\frac{1}{2 r_{e,\ell}^{2\ell-1} } k^2 \,. 
\label{eq:ert_lwave}
\end{equation}
The quantities $a_{\ell}$ and $r_{e,\ell}$ thus defined are known as the scattering length and the effective range, respectively. 
This approximation describes the phase shift with good accuracy  at sufficiently low energies. Consequently, if one partial wave dominates the scattering process, the velocity dependence of the cross section is  determined by only two parameters.

Let us focus on the $S$-wave case, which  dominates the low-energy scattering rate in many situations of interest. In this case, the cross section is given by\,\footnote{From now on, for simplicity we will omit the subscript $\ell=0$ for the scattering length and the effective range in the $S$-wave.} 
\begin{align}
\sigma_0=\frac{4\pi}{k^2} \sin^2 \delta_0 \approx \frac{4 \pi  a^2}{1+k^2 \left(a^2-a r_{e}\right)+\frac{1}{4} a^2 r_{e}^2 k^4}
\,.
\label{eq:swave_case}
\end{align}
Note that the unitarity bound, $4\pi/k^2$, is saturated for $|a|\to\infty$. 
One important example of this kind is the case of particles of mass $m$ interacting via the Yukawa potential 
\begin{equation}
V(r) =\pm\alpha \frac{e^{-m_\phi r}}{r}\,.
\label{eq:Yuk}
\end{equation}
 Fig.~\ref{fig:alpha}  compares the  numerically-evaluated cross section and the approximation  based on Eq.~\eqref{eq:swave_case} for a particular region of the parameter space.\footnote{Note that it is essential to go beyond the lowest-order perturbation theory (first Born approximation) $\sigma = 4\pi (\alpha m/m_{\phi}^{2})^{2}$, which does not depend on the sign of $\alpha$ nor produces spikes in the full calculation.  Effective range theory reproduces both correctly.} Despite its simplicity, the  effective range approximation works very well, \emph{i.e.} high-order terms of $k^2$ in Eq.~\eqref{eq:ert_lwave} can be neglected.  
In particular, it is able to reproduce the peak structure of the cross section. As is well known, such peaks are related  
to zero-energy bound states induced by the attractive potential.

A similar example is the non-relativistic scattering of two nucleons. In the case of proton-neutron collision, for the spin-one channel and kinetic energies up to a few MeV, Eq.~\eqref{eq:swave_case} accurately describes the velocity dependence of the corresponding cross section  with  $a = \unit[5.42]{fm}$ and $r_e=\unit[1.75]{fm}$~(see e.g.~\cite{Noyes:1973zd, Stoks:1993tb}).  This is also related to a bound state: the deuteron.

\begin{figure}[t]
\includegraphics[trim=0.cm 0cm 0.0cm 0.cm,clip,width=\textwidth]{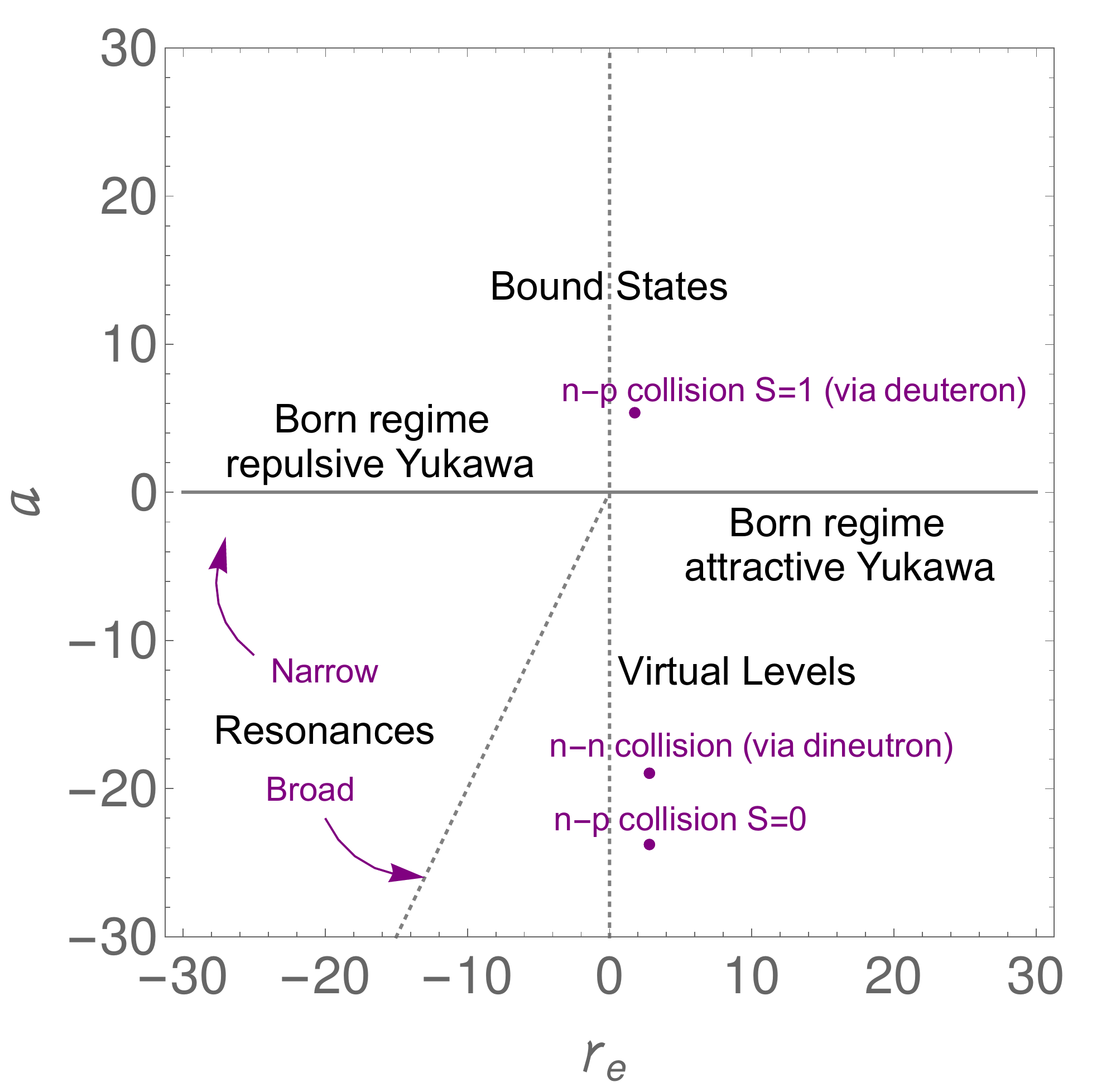}
\caption{
 Sketch of the scattering length and the effective range, which  determine the  cross section using Eq.~\eqref{eq:swave_case}.  They simultaneously parametrize the  non-relativistic scattering in seemingly different theories including  that induced by a Yukawa force, collisions via Breit-Wigner resonances, scatterings induced by bound state or virtual level, and the collision of non-relativistic protons (p) and neutrons (n).  For the last case, the displayed values in units of fm accurately describe the experimental data~\cite{Stoks:1993tb}. 
}
\label{fig:arplane}
\end{figure}

Likewise, the collision of two neutrons --for which the total spin is zero-- can be characterized by $a = \unit[-18.9]{fm}$ and $ r_e=\unit[2.75]{fm}$. In contrast, in this case  no real bound state exists. Instead, the scattering is induced by a virtual level,\footnote{A bound state is a pole in the scattering amplitude $f_{\ell}(k)$ along the positive imaginary axis on the complex $k$ plane, which corresponds to exponential damping of the radial wave function $\propto e^{i k r}$.  When parameters of the potential are varied, a pole may move to the negative imaginary axis, which no longer describes a bound state because the wave function grows exponentially.  However, the existence of a pole in the scattering amplitude can produce a pronounced enhancement in the cross section and is hence important.  In this case, the pole is called a virtual level.  It should not be confused with virtual particles or virtual states that refer to intermediate particles or states (propagators) in perturbation theory. } commonly known as the dineutron (see Fig.~\ref{fig:arplane}).

Equally interesting is the fact that  Breit-Wigner resonances can also be described using the effective-range approximation. For simplicity let us suppose that the colliding particles are scalars with  the same  mass $m$, and that the resonance has spin $\ell$. Thus, if the energy, $E=k^2/m$, is sufficiently close to the resonance $E_R$, the cross section is dominated by the partial wave $\ell$ so that 
\begin{equation}
\sigma_\ell= 
\frac{4\pi (2l+1)}{m E}\frac{\Gamma(E)^2/4}{\left(E-E_R\right)^2+\Gamma^2(E)/4}\,.
\end{equation}
 The width in general varies with the energy in such way that $\Gamma(E)\propto E^{(2\ell+1)/2}$\,(see e.g.~\cite{Chu:2018fzy}). Using the effective-range approximation to the phase shift (Eq.~\eqref{eq:ert_lwave}), we find that the  cross section, $\sigma_\ell=4\pi(2\ell+1)\sin^2\delta_\ell/k^2$,  exactly matches the previous formula with
\begin{align}\label{eq:EResonant}
	a_\ell= -\frac{\Gamma(E_R)^{1\over 2\ell+1} }{2^{1\over 2\ell+1} E_R^{2\ell+3\over 4\ell+2} m^{1\over 2}  } \,,~~ r_{e,\ell} = -{ 2^{2\over -2\ell+1} E_R^{2\ell+1 \over -4\ell+2} \over \Gamma^{1\over -2\ell+1} m^{1\over 2} }   \,.
\end{align}

Far from the resonance, some deviations are expected. In fact, as shown in right panel of Fig.~\ref{fig:alpha}, the effective-range approximation also fails close to the  antiresonances, {\it i.e.}\/,  where the cross section vanishes.  We will elaborate more on these cases in Section~\ref{sec:gen}. Likewise, when the range of the Yukawa potential, $m_\phi^{-1}$, is close to or larger than the de Broglie wavelength of the incoming particles, $k^{-1}$, the approximation fails. 
This region, usually referred to as classical regime, corresponds to $m_\phi\lesssim \unit[5]{MeV}$ for the parameter region of the right panel of Fig.~\ref{fig:alpha}. In fact, this is true for any potential,  for momenta larger than the inverse of the force range, not only must one include higher-order terms in Eq.~\eqref{eq:ert_lwave}, but also the differential cross section receives contributions from high partial waves.  In this case, the exact values of more phase shifts $\delta_\ell$ are needed to obtain the total scattering cross section. As a result, the effective-range approach can not be applied for long-range forces. For more details, see e.g.~\cite{Malley:1961fo, Berger:1965zz, vanHaeringen:1981pb}.

In summary,  the effective-range approximation properly describes  many  types of low-energy scattering. 
These cases differ in the magnitude and sign of the effective range parameters, as will be explained in Section~\ref{sec:models}  and as sketched in Fig.~\ref{fig:arplane}.

\section{Astrophysical implications}
\label{sec:SIDMv}

The main hypothesis of SIDM paradigm is that small-scale DM halos such as those of dwarf galaxies do not develop a high central density because
its DM particles self-scatter with a cross section per unit mass in the range $\unit[1-10]{cm^2/g}$~\cite{Vogelsberger:2012ku,Rocha:2012jg,Peter:2012jh}. 
On the other hand, observations of clusters of galaxies indicate that $\sigma/m\lesssim\unit[0.2]-\unit[1]{cm^2/g}$~\cite{Robertson:2016xjh,Randall:2007ph,Kaplinghat:2015aga, Bondarenko:2017rfu, 2018ApJ...853..109E, Harvey:2018uwf}.  Since in the former objects the average DM relative velocity is typically of the order of $10$ km/s, whereas in clusters of galaxies it is around $2000$ km/s, a velocity-dependent cross section is required in order to accommodate both.

Before discussing this in detail, let us note that we use $m$ for the DM mass, $v$ for the relative velocity between two initial DM particles in the centre-of-mass (CM) frame,   $m_\star = m/2$ for the reduced mass  and $k=m_\star v = mv/2$ for each incoming DM momentum.

\begin{figure}[tb]
\includegraphics[trim=0.cm 0cm 0.0cm 0cm,clip,height=0.305\textheight]{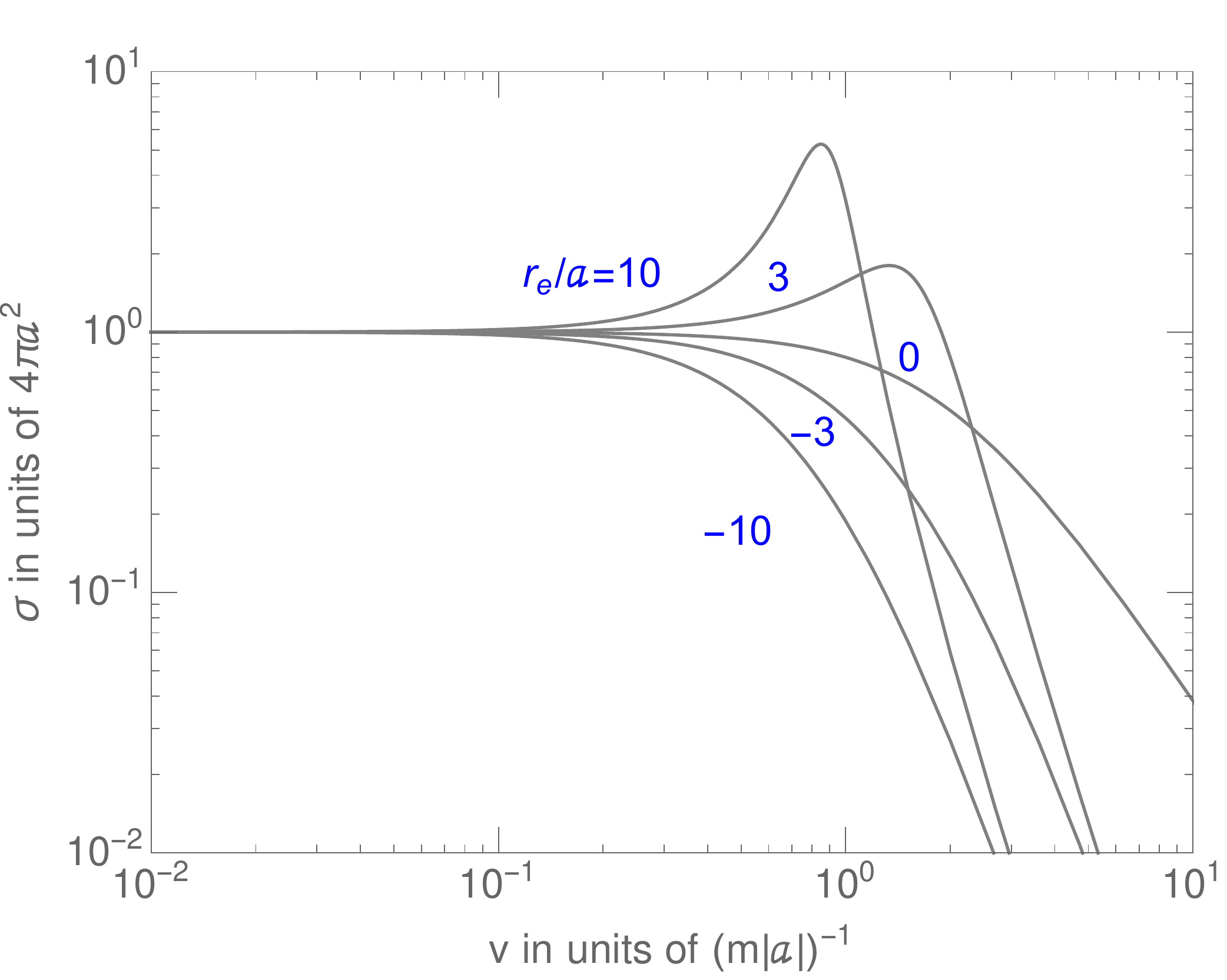}
\caption{ Self-scattering cross section as a function of the velocity for the indicated ratios of the effective range to scattering length.} \label{fig:sigmavsv}
\end{figure}

\begin{figure*}[t]
\includegraphics[trim=0.cm 0cm 0.0cm -1.cm,clip,height=0.345\textheight]{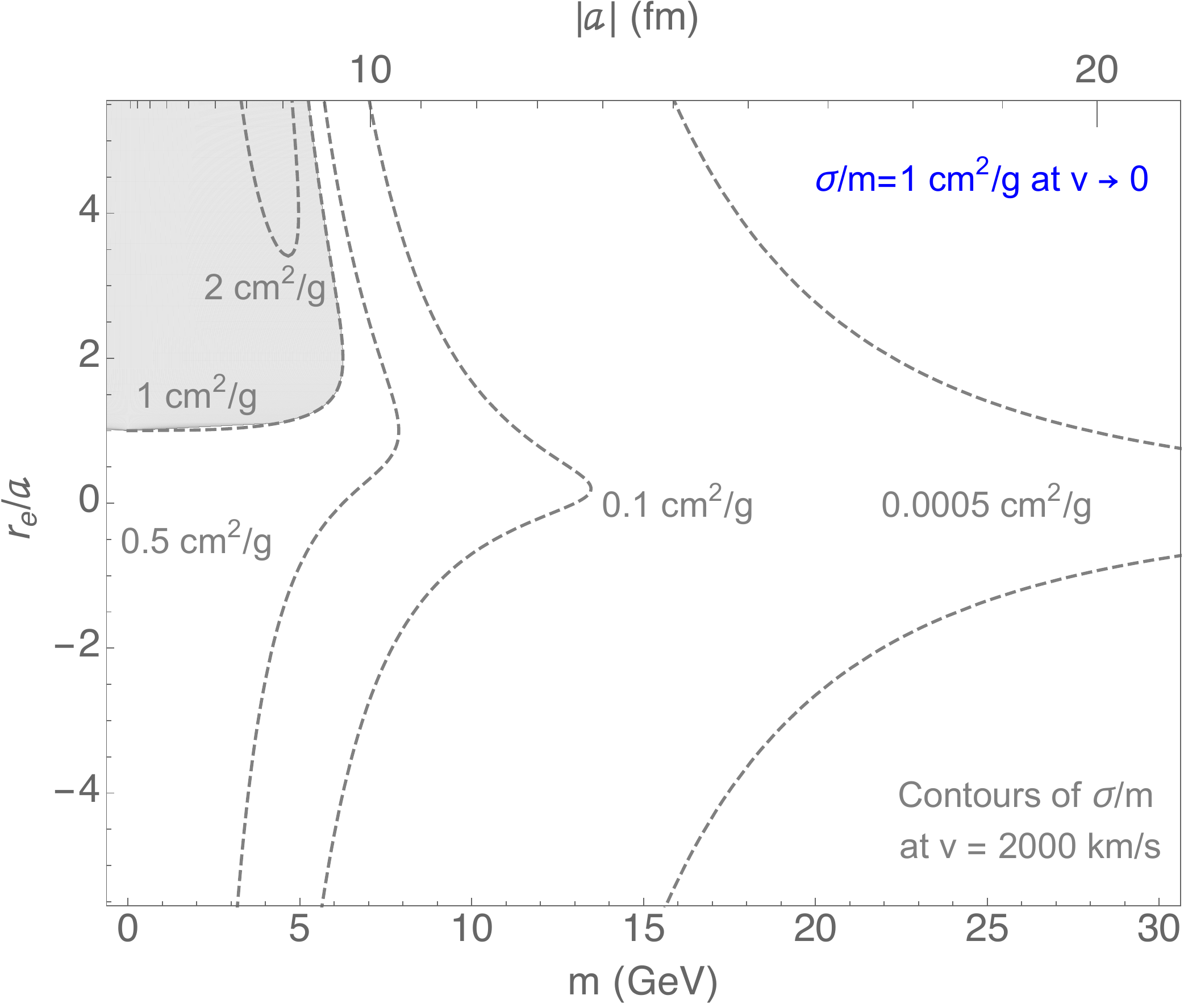}~
\includegraphics[trim=1.3cm 0cm 0.0cm -1.cm,clip,height=0.345\textheight]{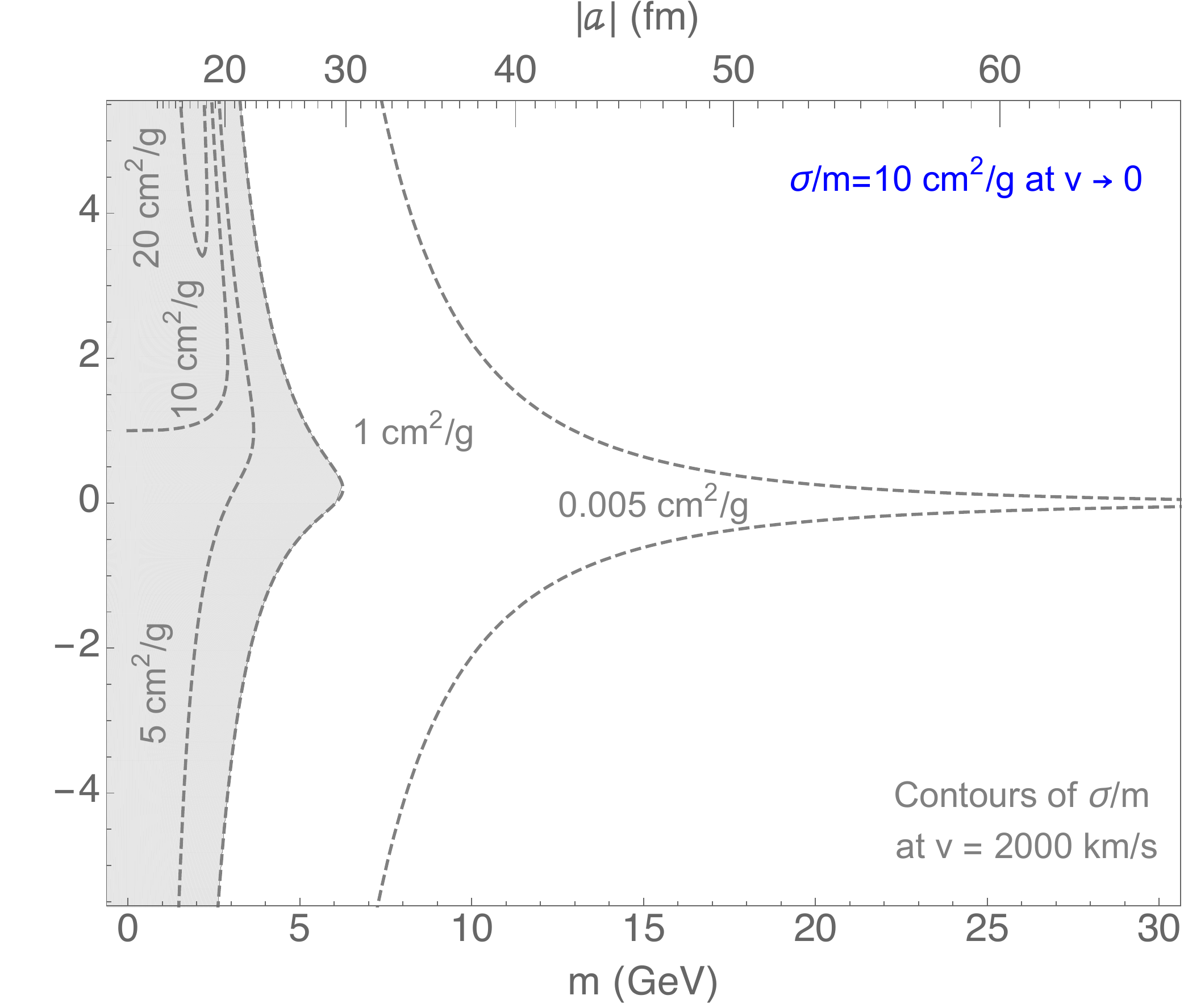}
\caption{Contours of the cross section per unit of mass at cluster scales ($v\sim\unit[2000]{km/s}$) for the indicated cross sections at zero velocity. In the parameter space shown in each plot, the latter coincides within 1\% with the cross section at dwarf-galaxy scales ($v\sim \unit[10]{km/s}$).  The gray area, where $\sigma/m\gtrsim\unit[1]{cm^2/g}$, is excluded by cluster observations~\cite{Robertson:2016xjh,Randall:2007ph,Kaplinghat:2015aga}.  }\label{fig:con}
\end{figure*}

\subsection{Velocity dependence in effective range theories}

In the effective-range framework, the velocity-dependent cross section is given by~\footnote{ The transfer cross section, $\sigma_T = \int d\sigma (1-\cos \theta)$, is typically used as a proxy for the scattering effects in DM halos. This is because, on the one hand,  $\sigma_T$  
takes into account that perpendicular scattering is most efficient for thermalizing the DM halo and affecting structure observables. On the other hand,  SIDM studies often discuss scatterings induced by the exchange of a light mediator, which exhibits a divergence in the forward direction  regularized by the transfer cross section. For  $S$-wave scattering, $\sigma_T$ and $\sigma_0$ coincide and are therefore  interchangeable.}
\begin{eqnarray}
{\sigma}(v) =  {4\pi a^2}\left(\left(1- \frac{1}{8}\frac{r_e}{a} (m av)^2\right)^2+\frac{1}{4} (m a v)^2 \right)^{-1}\,,
\label{eq:eqsigma}
\end{eqnarray}
where the signs of the scattering length and the effective range only enter in 
the equation via their ratio, and can not be separately constrained  by studying the velocity dependence of the DM scattering.

The velocity-dependence of the scattering cross section is also shown in Fig.~\ref{fig:sigmavsv}. At very low velocities the cross section is roughly constant and equal to $4\pi a^2$. If $r_e/a<1$,  the cross section monotonously decreases, most appreciably for high DM velocities, $v \gtrsim (m |a|)^{-1}$. In contrast, if $r_e/a>1$, the cross section increases with $v$ until it reaches the maximum  $4\pi r_e^2 a/(2r_e-a)$ at $v_\text{peak} = 2 (m |r_e|)^{-1} \sqrt{2(r_e/a-1)}$ and then  decreases.  If $|a| \ll 1/m$, the corresponding  cross section can be considered as a constant in all realistic DM halos.

To numerically illustrate this  for halos of various sizes,  Fig.~\ref{fig:con} shows the contours of the self-interaction cross section per unit mass at $v=\unit[2000]{km/s}$ for  $\sigma/m|_{v\to 0}$ equal to $\unit[1]{cm^2/g}$ (left) and $\unit[10]{cm^2/g}$ (right). We would like to note that  $\sigma/m|_{v\to 0}$ approximates the corresponding values in dwarf scales at  1\% level in the parameter space shown in the figure.  
From Fig.~\ref{fig:con}, we conclude that GeV SIDM is associated with scattering lengths of several fm and that sub-GeV SIDM is either excluded by cluster observations or requires a cross section  of around $\unit[1]{cm^2/g}$ throughout all scales of interest. While some of these conclusions have been obtained in specific SIDM scenarios  such as those involving a light mediator~\cite{Tulin:2013teo} or  resonant SIDM~\cite{Chu:2018fzy}, we would like to emphasize that these conclusions apply to any model where the effective-range approach applies.

\begin{figure*}
\includegraphics[trim=0.cm 0cm 0.0cm 0cm,clip,width=0.48\textwidth]{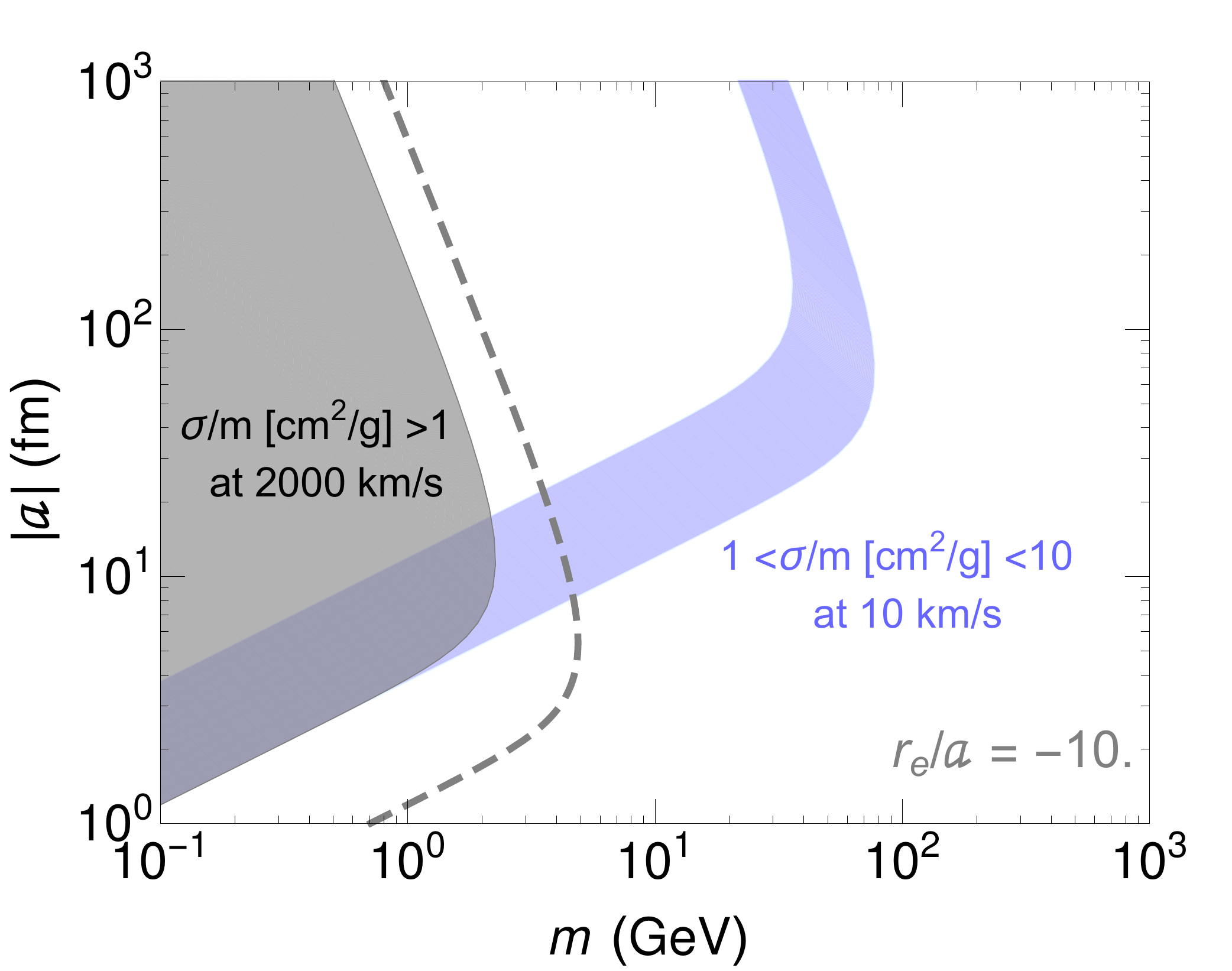}
\includegraphics[trim=0.cm 0cm 0.0cm 0cm,clip,width=0.48\textwidth]{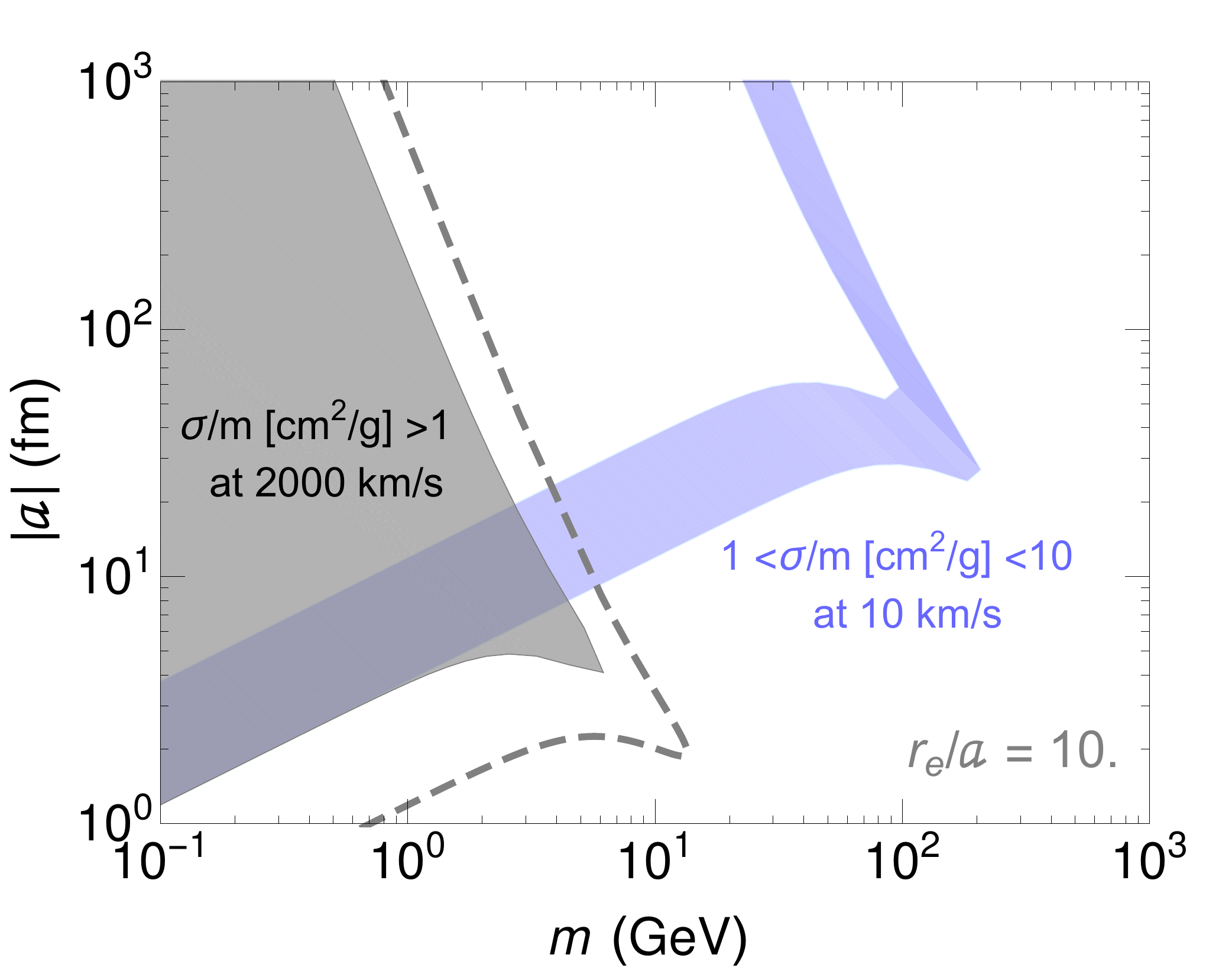}
\includegraphics[trim=0.cm 0cm 0.0cm 0cm,clip,width=0.48\textwidth]{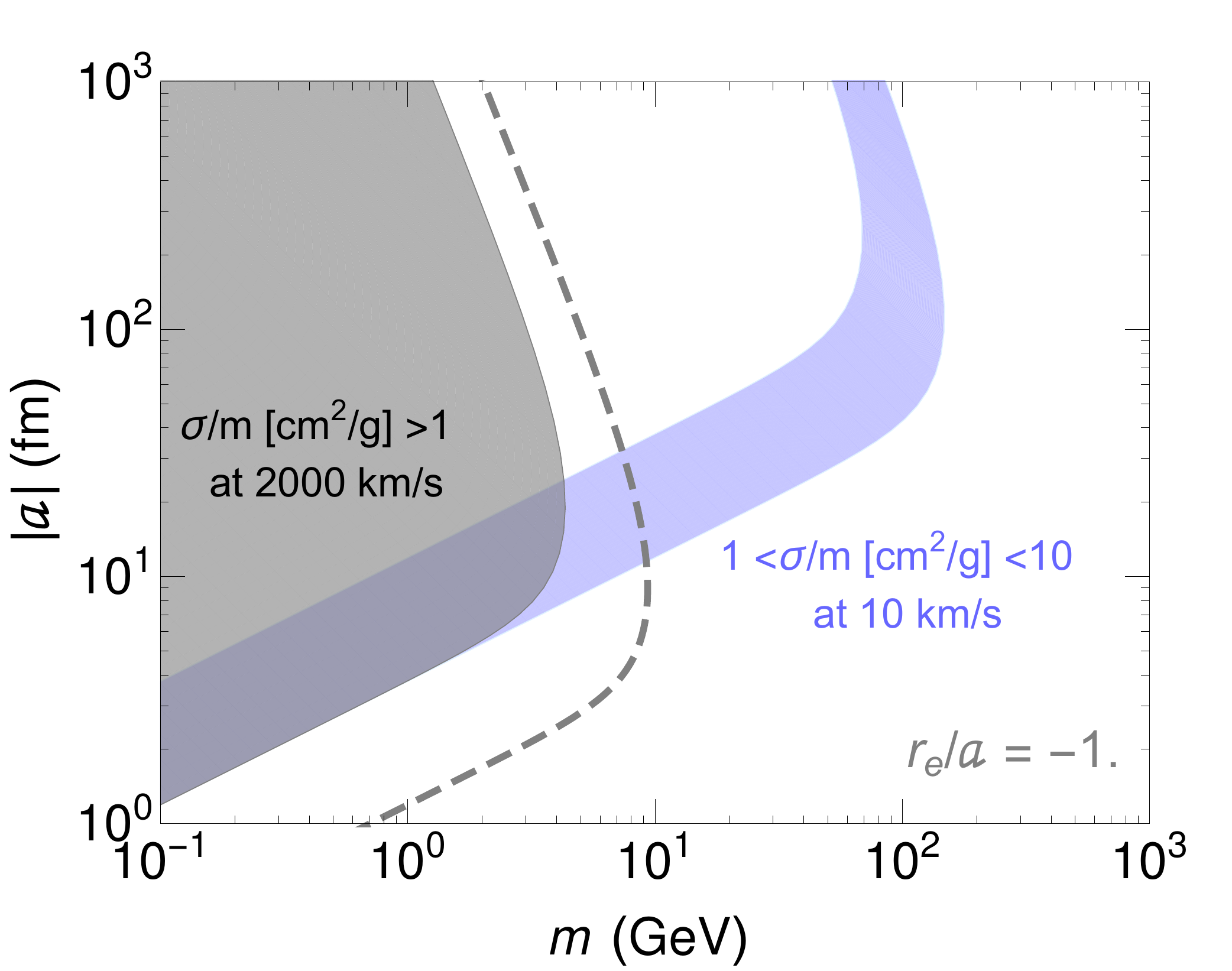}
\includegraphics[trim=0.cm 0cm 0.0cm 0cm,clip,width=0.48\textwidth]{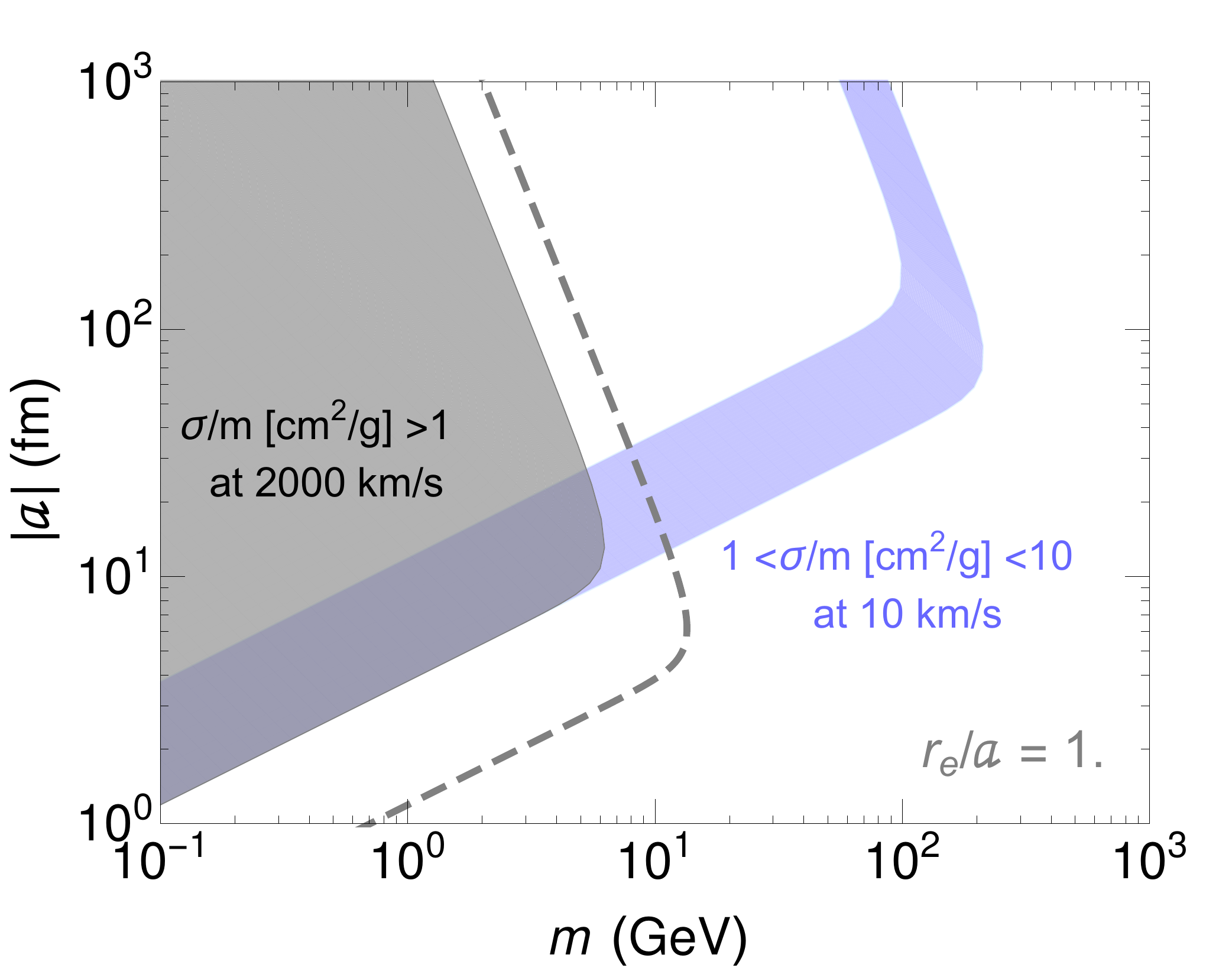}
\includegraphics[trim=0.cm 0cm 0.0cm 0cm,clip,width=0.48\textwidth]{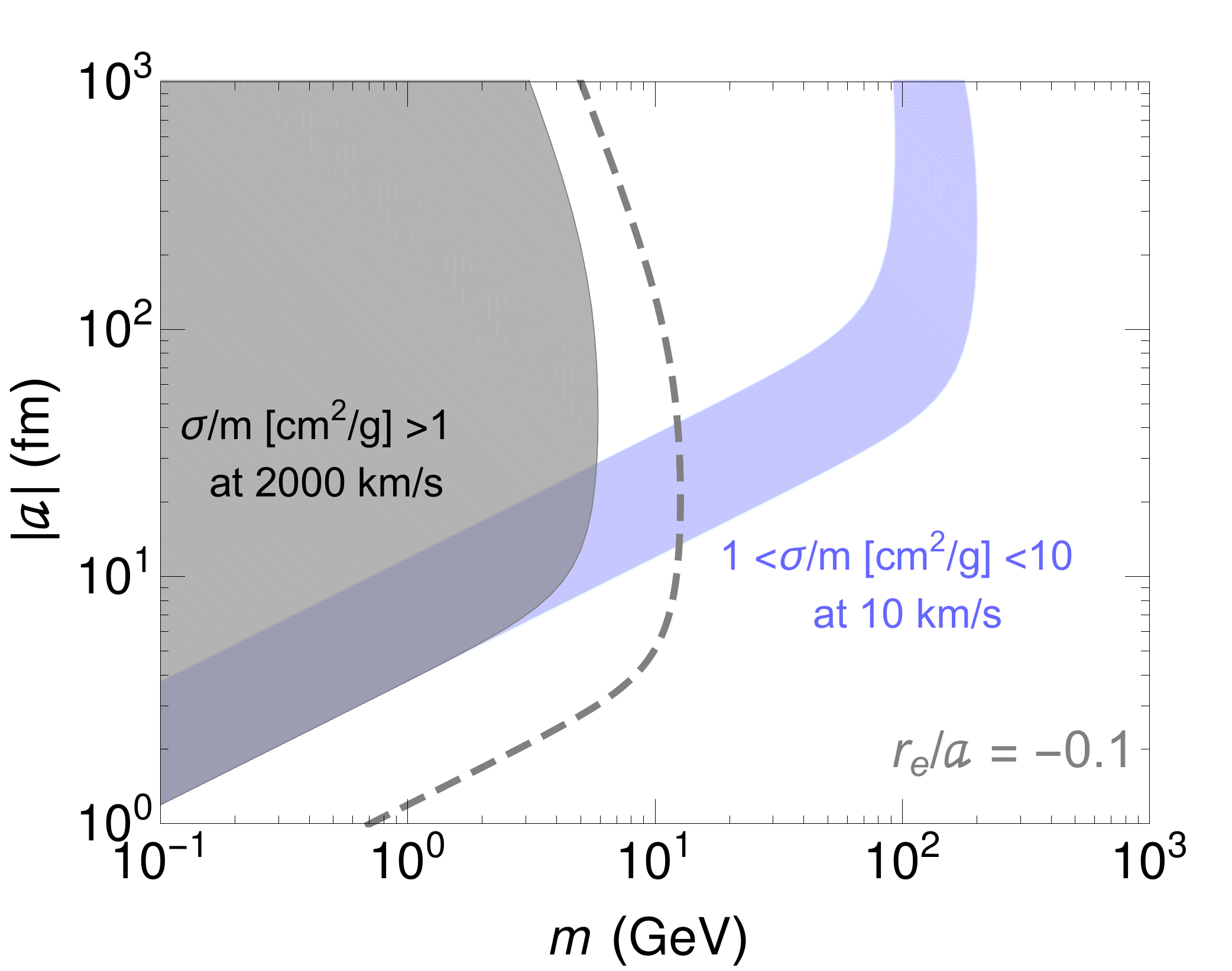}
\includegraphics[trim=0.cm 0cm 0.0cm 0cm,clip,width=0.48\textwidth]{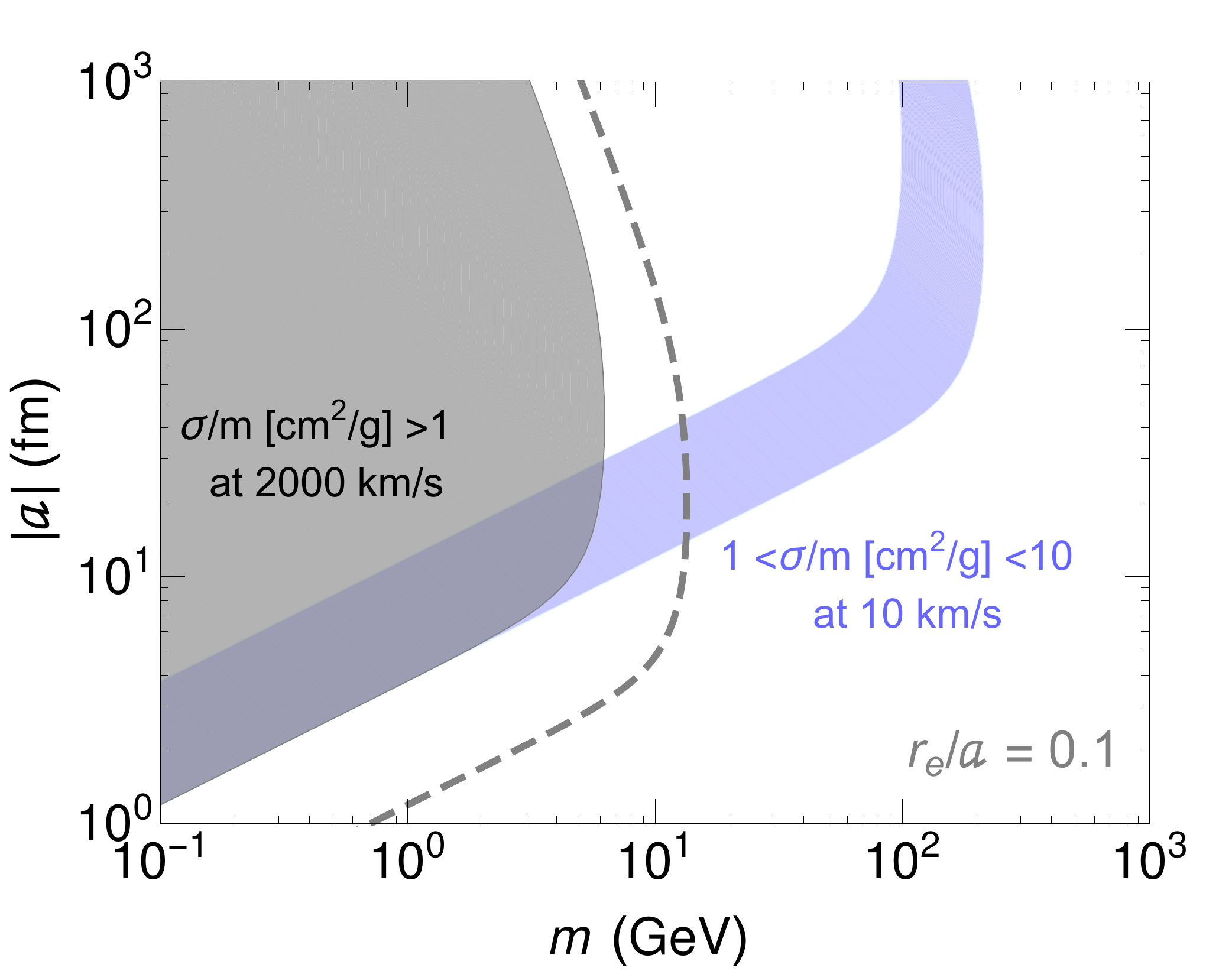}
\caption{Contours of  $\sigma/m$ within the range of $\unit[1]{cm^2/g}$--$\unit[10]{cm^2/g}$ at dwarf scales  (${v=\unit[10]{km/s}}$). The gray areas represent the exclusion limit from cluster-scale observables (${v=\unit[2000]{km/s}}$), and are extended to the gray dashed curves if one requires ${\sigma}/{m} \lesssim\unit[0.2]{cm^2/g}$ at cluster scales. }\label{fig:contours}
\end{figure*}

Fig.~\ref{fig:con} also suggests that the ratio $r_e/a$ is poorly constrained by the velocity dependence of the cross section. In the light of this and in order to consider a wider range of parameter space,   in Fig.~\ref{fig:contours}  we show three possibilities for the effective range compared to the scattering length. Concretely,  for each fixed value of $r_e/a$, we illustrate  the parameter space simultaneously satisfying
\begin{equation}
\begin{split}
 \unit[1]{cm^2/g}\lesssim\frac{\sigma}{m}\Bigg|_{v=\unit[10]{km/s}} &\lesssim\unit[10]{cm^2/g}\,,\\
 \frac{\sigma}{m}\Bigg|_{v=\unit[2000]{km/s}}&\lesssim \unit[1]{cm^2/g}\,.
\end{split}
\label{eq:contours}
\end{equation}
The former takes place within the colored region while the latter, as in Fig.~\ref{fig:con}, corresponds to the region not excluded by the gray area. Recent studies have claimed stronger constraints of $\unit[0.2]{cm^2/g}$ from observations of galaxy clusters~\cite{ Bondarenko:2017rfu, 2018ApJ...853..109E, Harvey:2018uwf}, which are indicated as a gray dashed line in each panel.

In each panel there are regions where all the constraints are simultaneously satisfied. This confirms our previous remark that the ratio $r_e/a$ is largely unconstrained.  Likewise, scenarios with a cross section of $\unit[1]{cm^2/g}$ at dwarf and cluster scales are those for which the borders of the gray and the colored regions lie on top of each other and  --as mentioned above-- they correspond to DM masses below a few GeV.

\subsection{Realistic velocity distributions}

So far we have assumed a monochromatic velocity distribution  for all DM particles in each halo. 
Below, we take into account the realistic distribution of DM velocities and then consider the corresponding average cross section for individual DM halos. The former is typically achieved by assuming a Maxwell-Boltzmann distribution with a cut-off scale:
\begin{equation}
f (v,v_0) = \frac{4 v^2e^{-{v^2}/{v_0^2}}}{\sqrt{\pi}v_0^3}\, \Theta \left(v_\text{max}-v\right) \,,
\label{eq:fBoltzmann}
\end{equation}
where $v_\text{max}$ is the escape velocity and $v_0$ is a parameter determining the typical velocities in the DM halo. For $v_\text{max}\gg v_0$, the average velocity is $\langle v \rangle =  2 v_0/ \sqrt{\pi}$.
More concretely, we take the average cross section as $\langle \sigma v\rangle/ \langle  v\rangle =  ({ \int^{\infty}_0 f(v,v_0)\, \sigma v \, dv})/({\int^{\infty}_0 f(v,v_0)\, v \, dv})$. 
Using a Maxwell-Boltzmann distribution would at most modify the curves of Fig.~\ref{fig:contours} mildly, so the conclusion of the previous subsection remains unchanged. For a  detailed calculation and a numerical comparison of $\langle \sigma v \rangle/\langle v\rangle $ and $\sigma (\langle v\rangle)  $,  see Appendix.~\ref{app:averageV}.

\begin{figure}[t]
\begin{tabular}{lllll} \hline 
  &   ~S1 &  ~S2 &  ~S3 &  ~S4 \\[0.05cm]
\hline 
$a$ (fm) &  19.2 &  25.6 &  3.8 &  37.4 \\ [0.05cm]
$r_e$ (fm) &  0.01 &  256.1 &  -57. &  -748.9 \\ [0.05cm]
$m$ (GeV) &  14.9 &  9.2 &  1. &  15.7 \\ [0.05cm]
\hline 
$ (m |a|)^{-1} ~$(km/s)&  205 &  250 &  15205 &  100 \\ [0.05cm]
$4\pi a^2/m($cm$^2$/g)  ~~~~~~ & 1.7 ~~~~~~ & 5. ~~~~~~ &  1.~~~~~~ & 6.3\\\hline 
\end{tabular}
\includegraphics[trim=0.0cm 0.0cm 0.0cm 0.0cm,clip,width=1\textwidth]{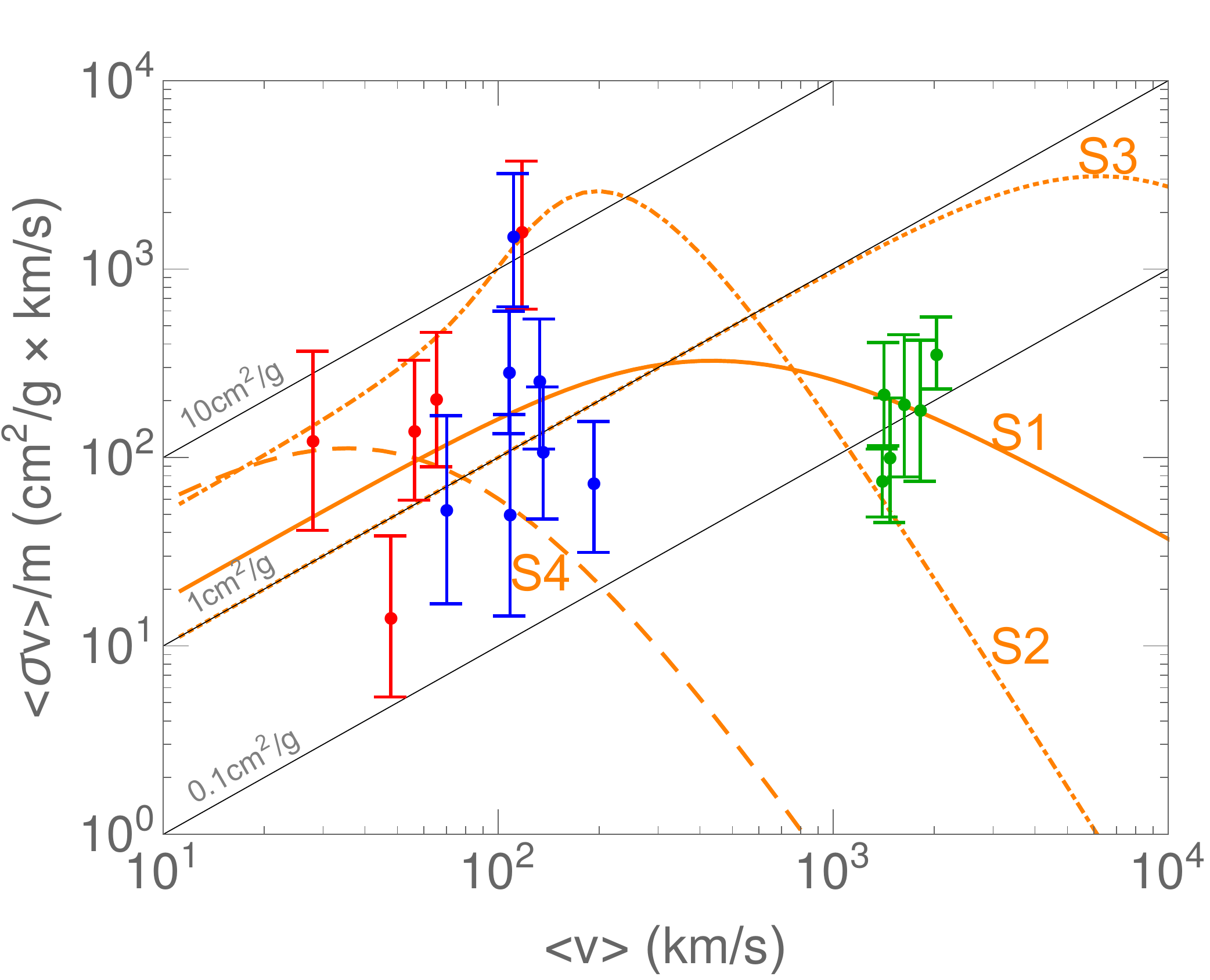}
\vspace{-.2cm}
\caption{Fit of DM self-interaction cross sections at various astrophysical scales using effective range approach. The points gives the inferred values of  $\langle \sigma v\rangle/m$ taken from \cite{Kaplinghat:2015aga}. The curves of S1-4 show the averaged $\langle \sigma v\rangle/m$ as a function of $v$, calculated for four benchmark parameter sets (see the top table). Note that although $a$ is set to be positive in the table, changing $a\to -a$ and $r_e \to -r_e$  simultaneously results in the same curve.  Among them, S1 gives the best-fit set.}
\label{fig:fit}
\end{figure}

On the observational side, extracting cross sections from experimental data is   challenging and  generally requires delicate N-body simulations at present.  An intermediate method is given by the  semi-analytical method  proposed in \cite{Kaplinghat:2015aga}, which allows to infer the velocity-averaged cross section per unit mass, $\langle \sigma v\rangle/m$, for a given DM halo. This method was applied  to five clusters from \cite{Newman:2012nw},  seven low-surface-brightness (LSB) spiral galaxies  in \cite{KuziodeNaray:2007qi} and six  dwarf galaxies of the THINGS sample~\cite{Oh:2010ea} (see also \cite{Valli:2017ktb}). Fig.~\ref{fig:fit} shows their resulting values in green,  blue and red, respectively.
 
Using the velocity-averaged scattering cross section, we can fit $|a|$, $r_e/a$ and $m$ to  the aforementioned  semi-analytical results and thus constrain the parameters of the effective-range theory. The best-fit point is shown in Fig.~\ref{fig:fit}  and corresponds to the benchmark S1.\footnote{As mentioned before, the sign of $a$  or $r_e$ can not be fit by studying the velocity dependence of the scattering. Moreover, since S1 has very small value of $r_e$, flipping the sign of either of them, in practice, gives the same cross section as a function of the velocity.}  As expected, it fulfills the condition stated in Eqs.~\eqref{eq:contours}. 

We would like 
to emphasize that the  points shown in Fig.~\ref{fig:fit} should be taken with caution, as  subtle effects, such as tidal stripping, still need to be further studied to understand the (sub-)halo dynamics (see e.g.  \cite{Sokolenko:2018noz, Kummer:2019yrb, Kaplinghat:2019svz, Sameie:2019zfo, Kahlhoefer:2019oyt} for recent discussions). In fact,  the need of a sizeable DM self-interaction at cluster scales is under  debate as this relies on the assumption that one can
robustly infer the existence of cores in clusters of galaxies. 
This motivates us to also consider other possibilities which are not necessarily fitting the green points but in agreement with  conservative bounds at cluster scales, $\sigma/m\lesssim\unit[1]{cm^2/g}$. These are the benchmarks S2,  S3 and S4 labeled in Fig.~\ref{fig:fit}.

The benchmark  S2  fits the dwarf and LSB data points fairly well with a relatively high cross section but is too low to fully accommodate the cluster points. This is because its peak is very pronounced with $r_e/a\gg 1 $  and  $\langle\sigma v\rangle$ decreases very rapidly for $\langle v\rangle $ greater than the peak velocity. This can be achieved with a narrow resonance~\cite{Chu:2018fzy}. 
 In contrast, another benchmark S3 describes an almost constant self-interaction cross section. Interestingly,  this is the benchmark that gives the lowest DM mass. As mentioned below, this is the sort of points expected in particle models with contact interaction or heavy mediators.   Finally, benchmark S4 describes a velocity-averaged cross section whose peak velocity is around 50 km/s and avoids potentially stringent  bounds on self-interaction cross section from massive galaxy/cluster observations.

\section{Interpreting $a$ and $r_e$ in terms of model parameters}
\label{sec:models}

In this section, we  discuss how the scattering length and the effective range are related to the model parameters of  several SIDM scenarios, including those with a light mediator~\cite{Spergel:1999mh, Feng:2009hw},  resonant SIDM~\cite{Chu:2018fzy}, as well as Strongly Interactive Massive Particles (SIMP)~\cite{Dolgov:1980uu, Carlson:1992fn, Hochberg:2014dra}. 

\subsection{Contact interaction}

The simplest model discussed for SIDM is
\begin{equation}
	V = \frac{1}{2} m^{2} \phi^{2} + \frac{1}{4!} \lambda \phi^{4}
\end{equation}
where $m$ is the DM mass and $\lambda$ the coupling.  It leads to
a constant cross section (within the Born approximation)
\begin{equation}
	\sigma_{0} = \frac{\lambda^{2}}{128\pi m^{2}}\ .
\end{equation}
In the effective-range framework, this happens for $|r_e|\ll |a|\ll k^{-1} $ so that 
\begin{equation}
\sigma_0 = 4\pi a^2.
\end{equation}
For the Born amplitude to be trusted, we need $\lambda \lesssim 1$, and hence
\begin{equation}
	m \lesssim 8.14\,{\rm MeV} ~\lambda^{2/3} 
		\left( \frac{1~{\rm cm}^{2}/{\rm g}}{\sigma_{0}/m} \right)^{1/3}.
\end{equation}
If we believe in the upper limit from the clusters, such contact interaction provides a poor fit to the data.  Note that the benchmark point S3 corresponds to a large coupling $\lambda \sim  10^{5}$ for 15\,GeV DM, where we can no longer trust the Born approximation.  Typically, the $\phi^{4}$ theory requires a UV completion of a strongly-coupled dynamics among dark matter particles.

\begin{figure*}[t]
\includegraphics[trim=0.cm 0cm 0.0cm 0.cm,clip,width=0.49\textwidth]{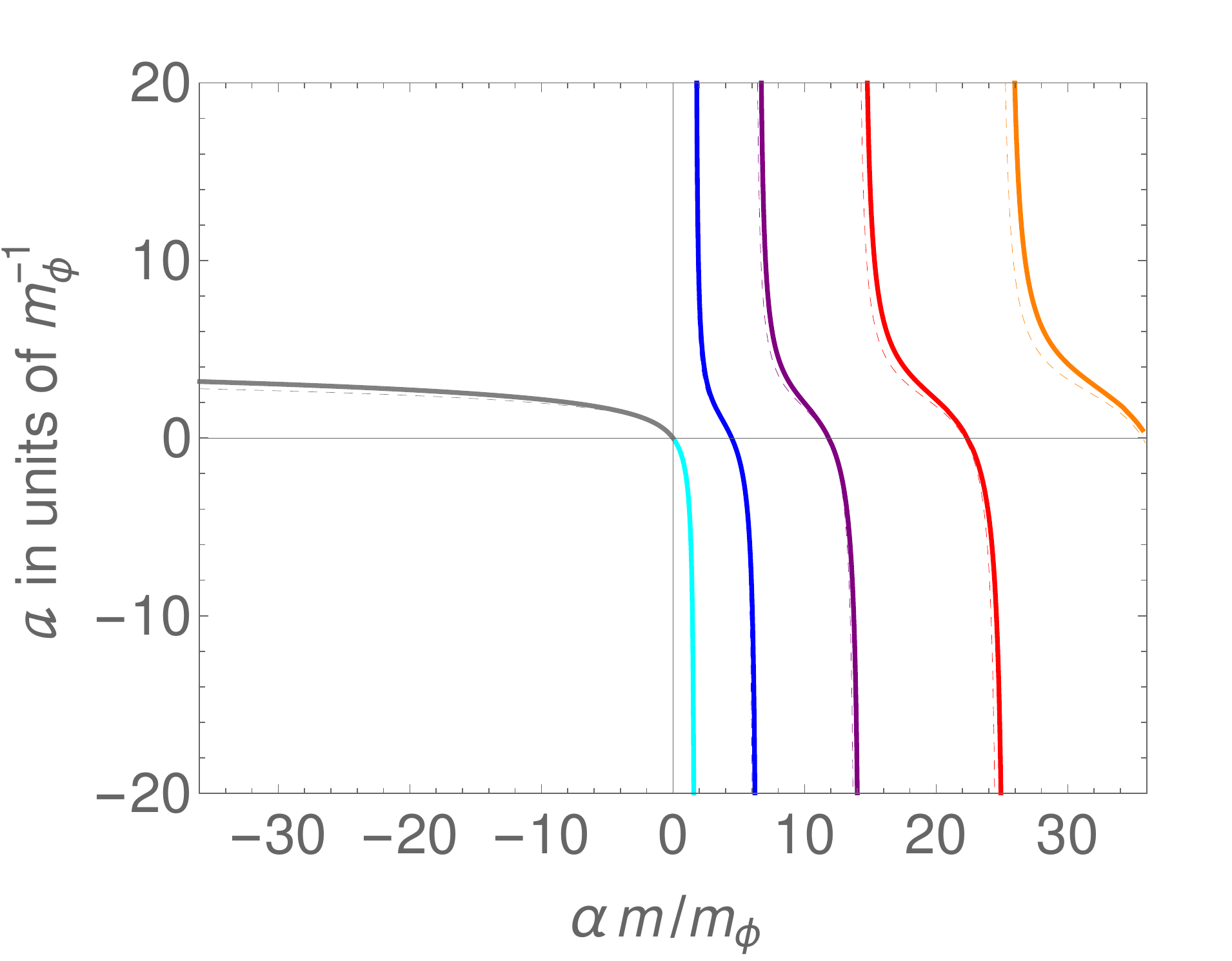}
\includegraphics[trim=0.cm 0cm 0.0cm 0.cm,clip,width=0.49\textwidth]{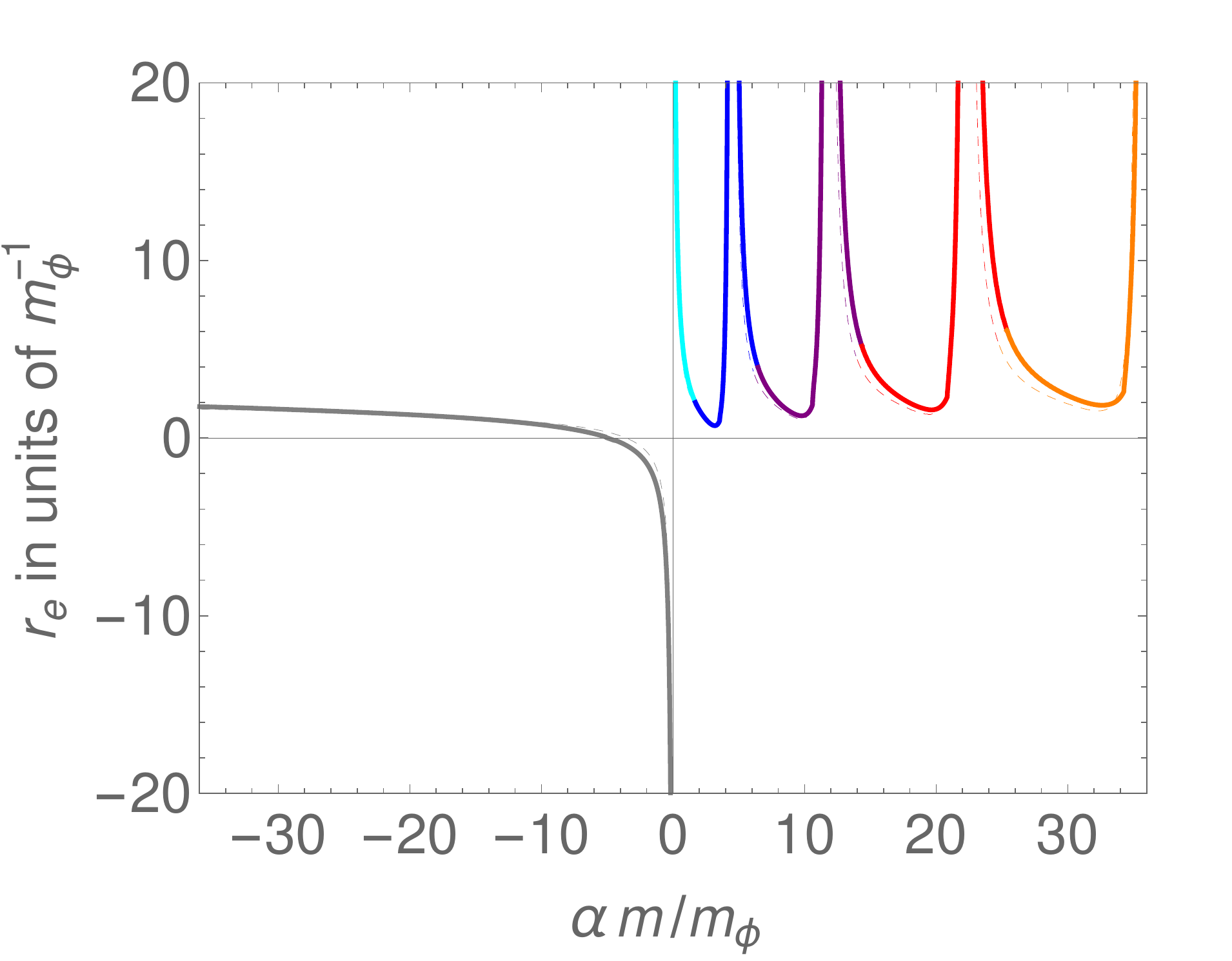}
\caption{
\emph{Left:} $S$-wave scattering length as a function of $\alpha m/m_\phi$ for the Yukawa (solid) and the Hulth\'en  (dashed) potentials. The case of a repulsive force ($\alpha<0$) is shown in gray. For the attractive case ($\alpha>0$), as $\alpha m/m_\phi$ increases,   a different color is chosen after the phase shift reaches an odd multiple of $\pi/2$. This indicates a \emph{parametric} resonance, where the cross section reaches a maximum (the unitarity limit) and the scattering length diverges. The antiresonances, $\delta=0$, correspond to vanishing scattering lengths and thus zero cross sections. \emph{Right:} Same as the left panel but for the $S$-wave effective range.}
\label{fig:Yukawa}
\end{figure*}

\begin{figure}[b]
\includegraphics[trim=0.cm 0.0cm 0.0cm 0.cm,clip,width=\textwidth]{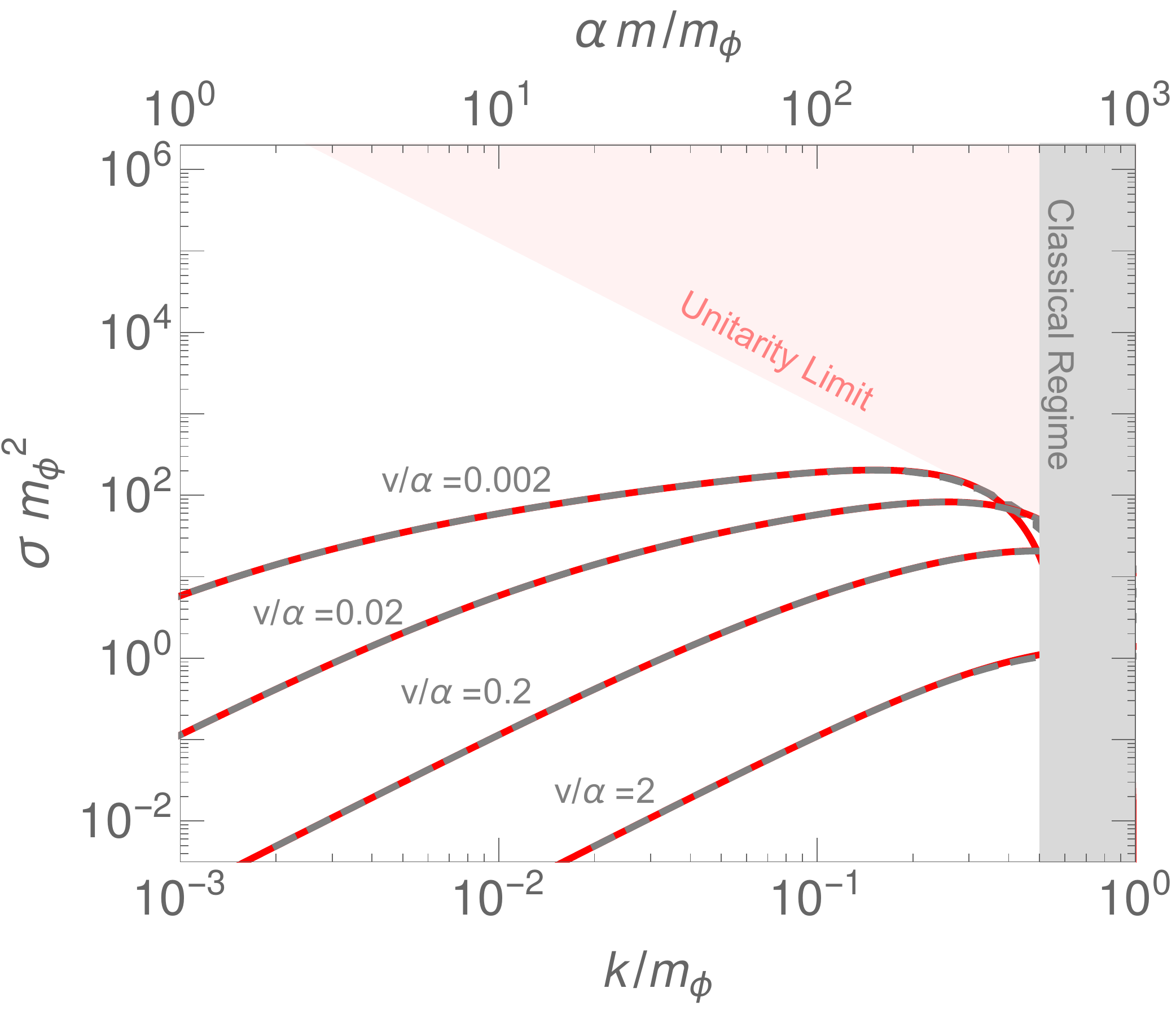}
\caption{$S$-wave scattering cross section for the repulsive Yukawa potential in Eq.~\eqref{eq:Yuk}. The red  solid lines are the numerical results, while the dashed  gray lines are given by the corresponding  effective-range approximation.}
\label{fig:approxRepul}
\end{figure}

\begin{figure*}[t]
\includegraphics[trim=0.cm 0.0cm 0.0cm 0.cm,clip,height=0.42\textwidth]{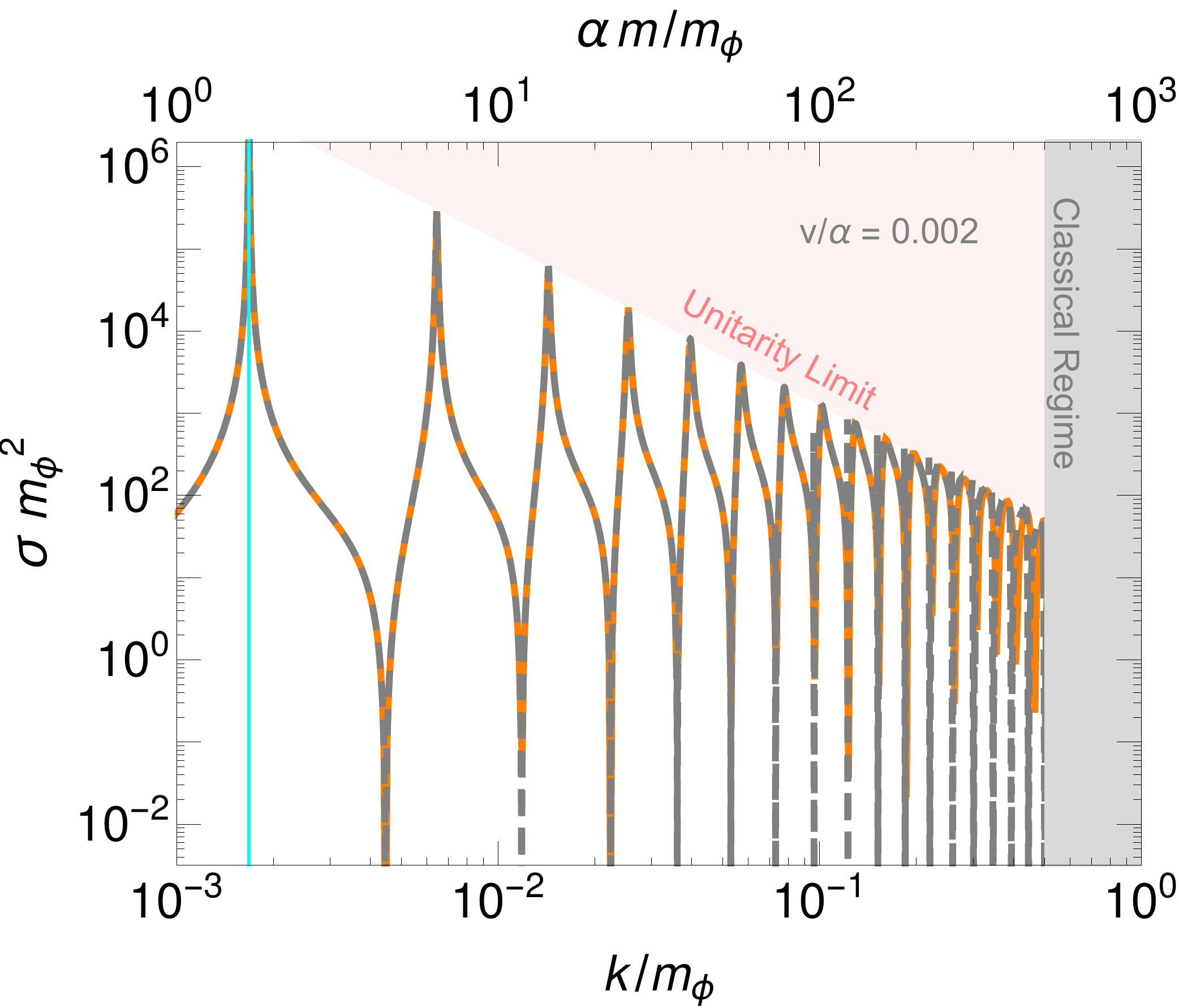}
\includegraphics[trim=1.2cm 0.0cm 0.0cm 0.cm,clip,height=0.42\textwidth]{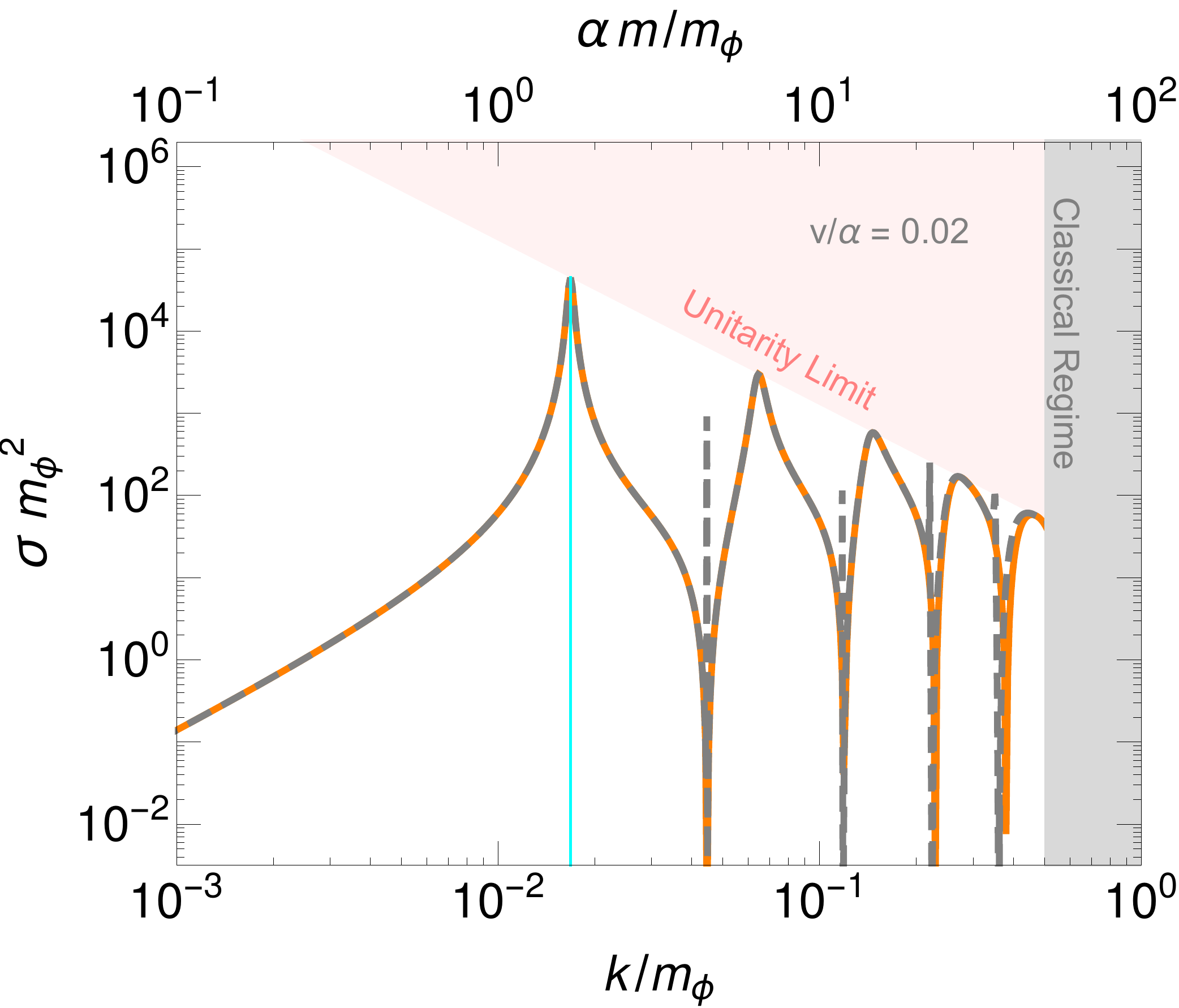}\\
\includegraphics[trim=0.cm 0.0cm 0.0cm 0.cm,clip,height=0.42\textwidth]{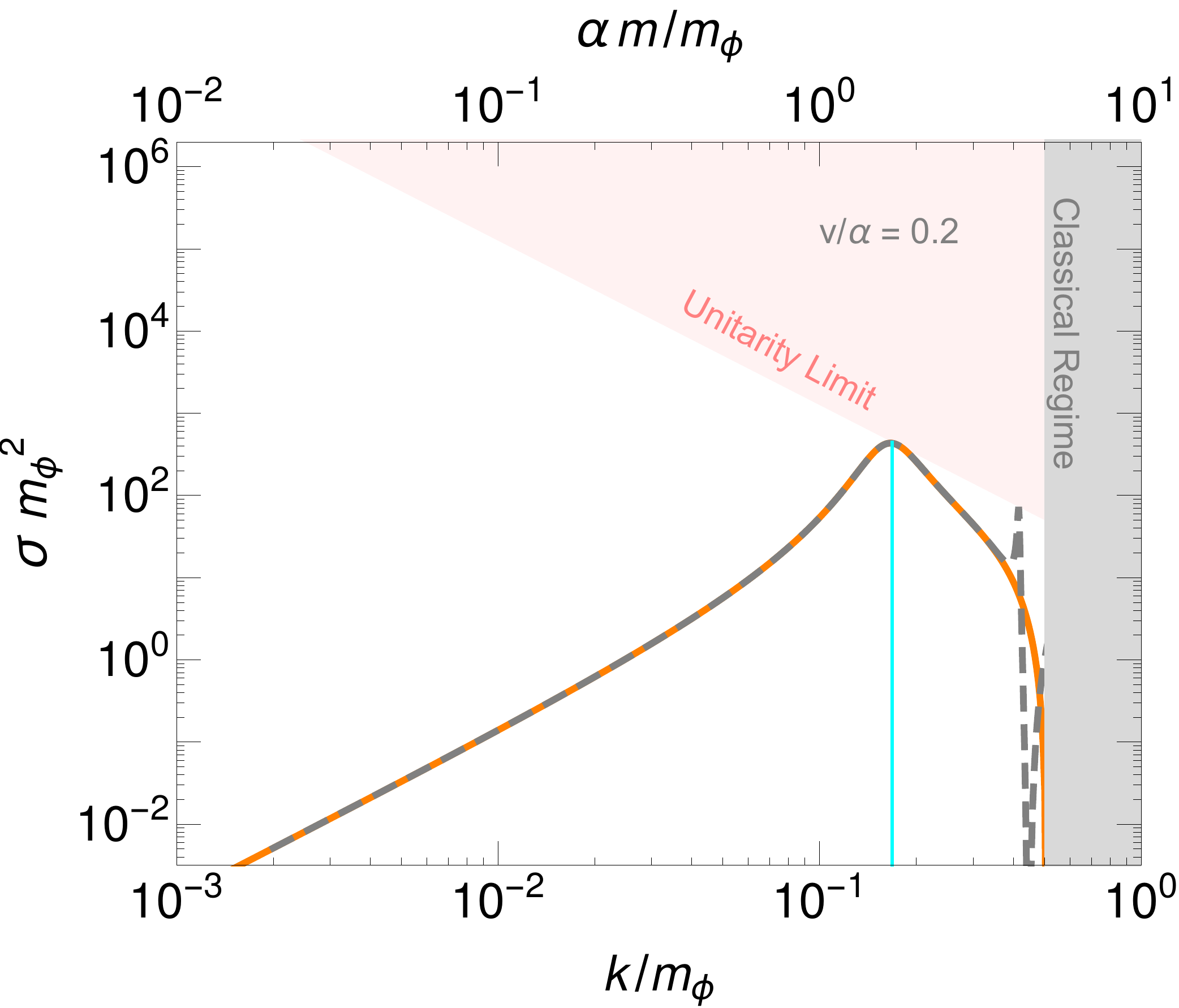}
\includegraphics[trim=1.2cm 0.0cm 0.0cm 0.cm,clip,height=0.4\textwidth]{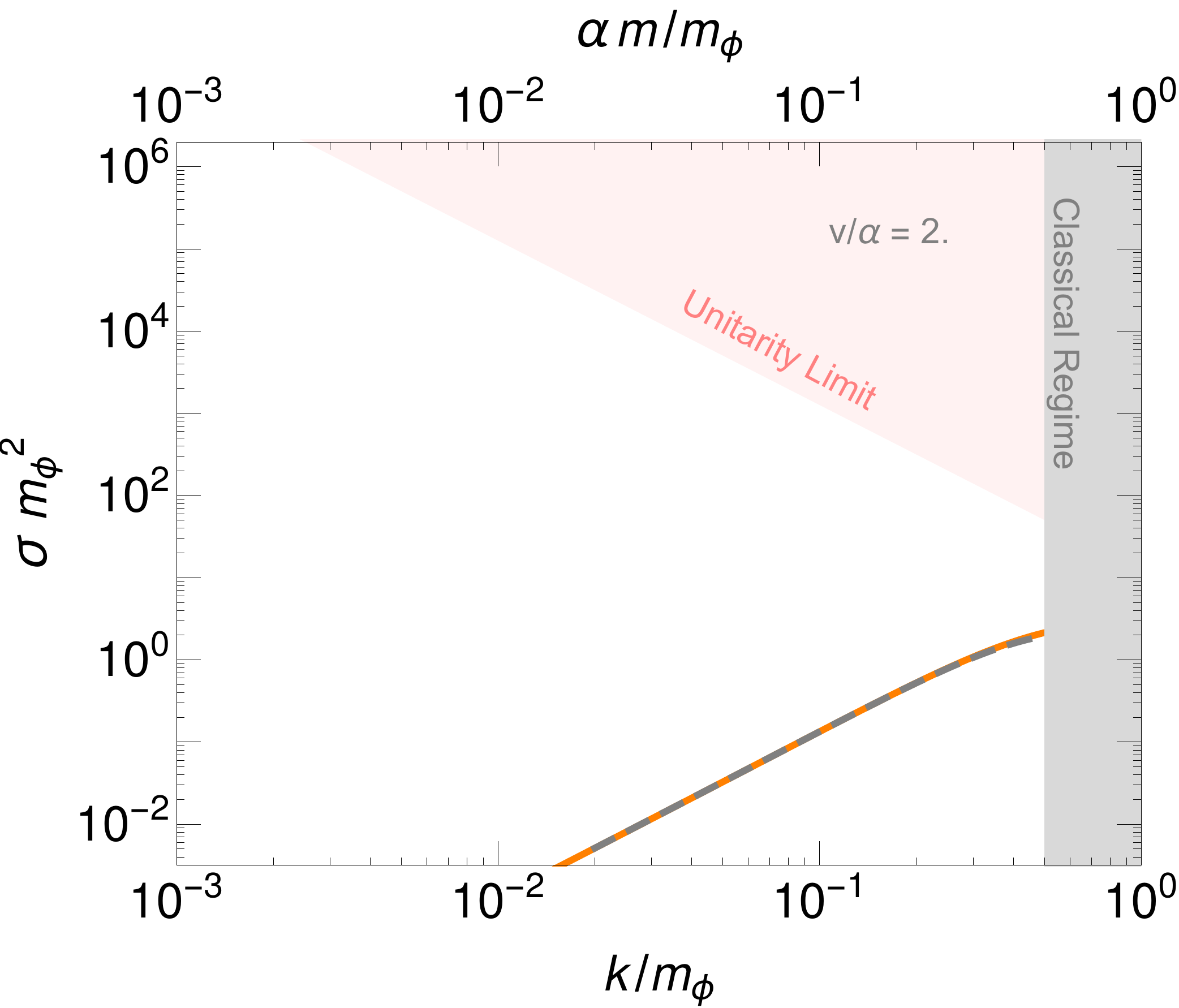}\\
\caption{
$S$-wave scattering cross section for the attractive Yukawa potential in Eq.~\eqref{eq:Yuk}. The orange  solid lines are the numerical results, while the dashed  gray lines are the corresponding  effective-range approximation. Vertical cyan lines correspond to the first resonance at $\alpha m/m_\phi \simeq 1.68$.}
\label{fig:approx}
\end{figure*}

\subsection{SIDM with a light mediator}
\label{sec:Yukawa}

Models in which non-relativistic DM is coupled to a boson of mass $m_\phi$ predict DM self-interactions mediated by the Yukawa potential of Eq.~\eqref{eq:Yuk}.  Using the numerical method discussed in Appendix~\ref{app:ERT}, we calculate the  $S$-wave scattering length and effective range together with the corresponding exact and approximated cross sections, for both repulsive and attractive cases.  The results are shown in Fig.~\ref{fig:Yukawa}, Fig.~\ref{fig:approxRepul} and Fig.~\ref{fig:approx}, respectively.
Notice that fixing $\alpha m/m_\phi$ and $v/\alpha$ determines all quantities in units of $m_{\phi}^{-1}$. 
 The figure also gives the ($a$, $r_e$) parameters of the Hulth\'{e}n potential $V(r)=\pm \alpha \delta e^{-\delta r}/(1-e^{-\delta r})$, which has been used to approximate the Yukawa potential by setting $\delta=\sqrt{2\zeta(3)}  m_\phi$~\cite{Tulin:2013teo}. Both potentials give similar effective-range parameters, and thus similar self-interacting cross sections.

\paragraph{The Born regime:} In this case $\alpha \,m \ll m_\phi$ and the phase shift can be found by solving the Schr\"odinger equation perturbatively. In this way,  according to Eq.~\eqref{eq:Born}, we have
\begin{eqnarray} \label{eq:YukawaB}
	\tan\delta_0 &\simeq& -m k \int^\infty_0 r^2 V(r) {\sin^2(kr) \over (kr)^2} dr \notag \\
	&=&   {m   \alpha k \over m_\phi^2 }\left(1- {2k^2 \over m_\phi^2} +{\mathcal O} \left( {k^4 \over m_\phi^4}\right) \right),\, 
\end{eqnarray}
which implies 
\begin{align}\label{eq:ERTB}
	a= -{\frac{ m \alpha}{m_\phi^2} }\,,\text{~~and~~~} r_e = {4\over   m  \alpha} \,.
\end{align}
 Therefore, in the limit of  very small $\alpha$, the scattering length is negligible,  the effective range $r_e$ is large and they have opposite signs. This behaviour is  clearly shown in Fig.~\ref{fig:Yukawa}.  As can be seen from the $k^2$ expansion of Eq.~\eqref{eq:YukawaB}, even in this Born regime, the effective range formula $k\cot \delta_0 = -1/a+r_ek^2/2$   can only approximate the $S$-wave phase shift for $k\ll m_\phi$.
The opposite case is the classical regime mentioned before, where higher partial waves have to be taken into account. 

\paragraph{The resonant regime:} 

Now we turn to the parameter regime satisfying $\alpha m \gtrsim  m_\phi$, where non-perturbative effects play an important role. In particular, the attractive case exhibits a very rich phenomenology. This is in sharp contrast to the repulsive case shown in Fig.~\ref{fig:approxRepul}, where the scattering cross section simply increase with larger couplings.

For attractive interactions, Fig.~\ref{fig:Yukawa} shows that as $m\alpha/m_\phi$ gradually  increases,  a critical value is reached at which the scattering length goes to negative infinity. This  corresponds to a phase shift approaching  $\pi/2$ from below and a cross section of $\sigma_0 = 4\pi/k^2$. Notice that this is the maximum value allowed by unitarity. Immediately after $m\alpha/m_\phi$ exceeds such a critical value,  $a$ becomes positively infinite.

Then,  with even larger $m\alpha/m_\phi$, the scattering length starts to decrease,  until it reaches zero, corresponding to a phase shift of $\pi$. This is the so-called antiresonance, where the cross section takes its minimum value. Further increasing $m\alpha/m_\phi$ leads to negative values for $a$, which eventually approaches negative infinity again. The same cycle repeats itself indefinitely.  Notice that this behavior of the cross section is responsible for the peak structure observed in Figs.~\ref{fig:alpha} and ~\ref{fig:approx}.
As mentioned in the previous section, there is a close connection between those peaks, where $|a|\to \infty$,  and the bound states that are formed due to the Yukawa potential.  
Below, we discuss this and how they are related to the poles of the scattering amplitude.

\begin{figure*}[t]
\includegraphics[trim=0.cm 0cm 0.0cm 0.cm,clip,width=0.48\textwidth]{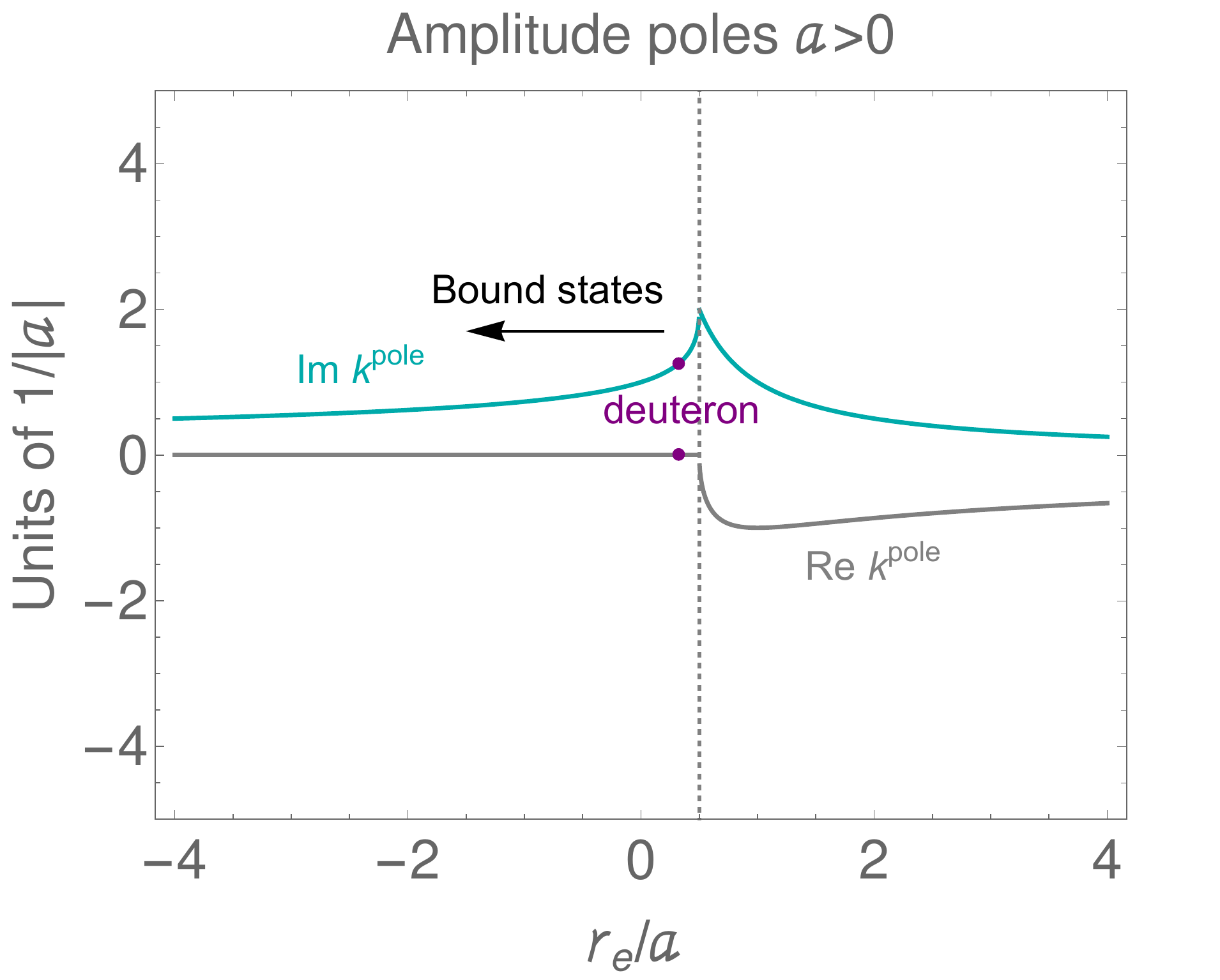}
\includegraphics[trim=0.cm 0cm 0.0cm 0cm,clip,width=0.48\textwidth]{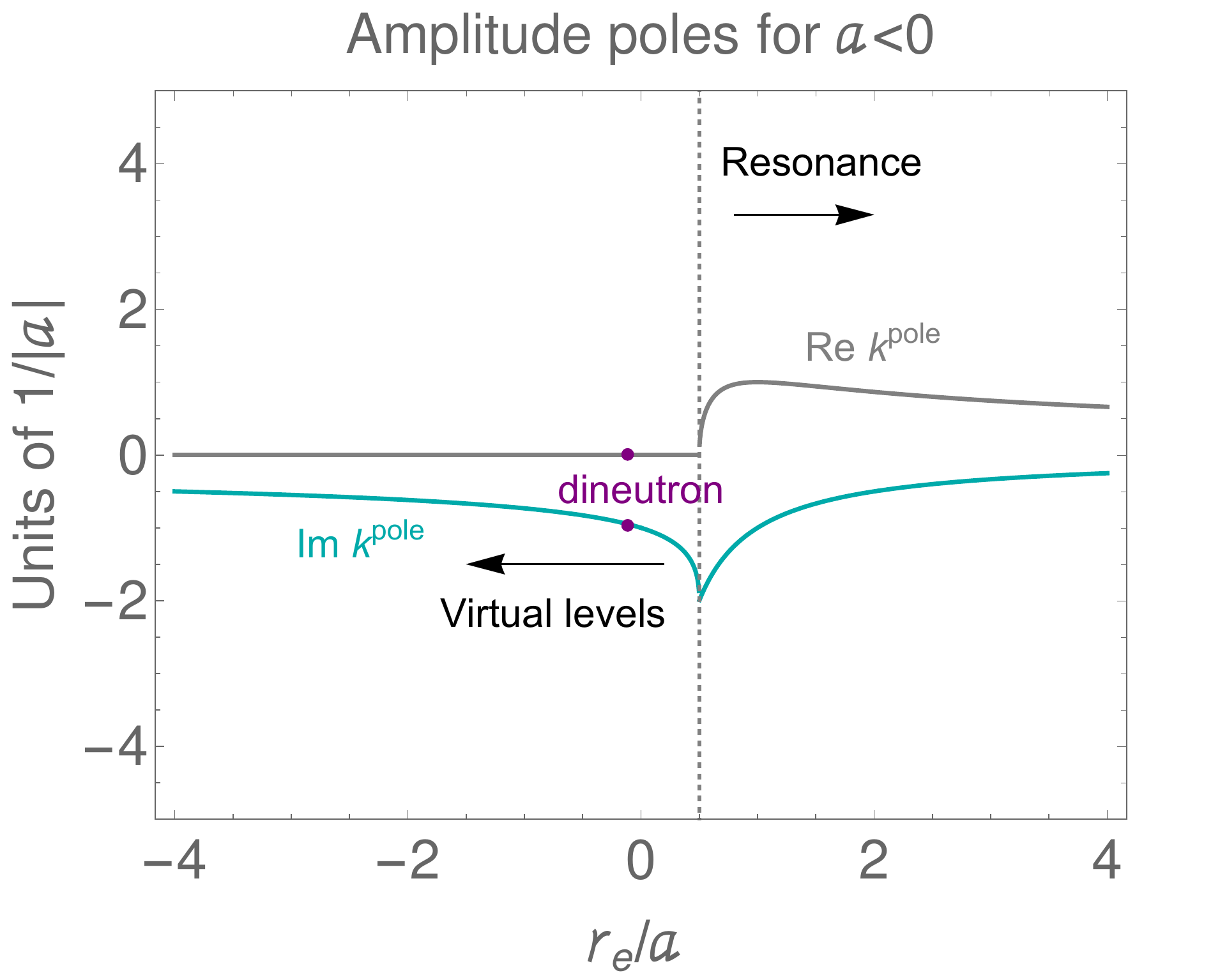}
\caption{The real and imaginary parts of $k^\text{pole}_{+}$ in units of $1/|a|$ as a function of $r_e/a$ for both $a>0$ (left) and $a<0$ (right). The corresponding physical states are labelled in texts. Note that in left panel the pole at $r_e/a > 1/2$ is unphysical (see footnote~\ref{fn:poles}).}
\label{fig:peak}
\end{figure*}

\subsection{SIDM via bound states or virtual levels}
\label{sec:Poles}

Eq.~\eqref{eq:swave_case} has the following poles 
\begin{align} 
 		k^\text{pole}_{\pm} = {i   \over a } \,{2  \over  1  \pm  \sqrt{ 1- {2r_e /a}  }} \,.
\label{eq:kpole}
\end{align}
Even though the poles are in general complex, they can influence the low-energy scattering if they are sufficiently close to the incoming particle momentum.  In fact, a close inspection of the  Schr\"odinger equation allows us to interpret them in terms of physical states.\footnote{For a textbook review of these topics,  see {Landau\,\&\,Lifshitz}~\cite{Landau:1991wop}.} In Fig.~\ref{fig:peak} we plot the real and the imaginary parts of $k^\text{pole}_+$, which is the closer pole to the real axis. 

For simplicity, let us consider first the case of a pure imaginary $k^\text{pole}$. The corresponding energy $E=(k^\text{pole})^2/(2m_\star)$ is negative, indicating  the existence of a bound state with binding energy $\epsilon=-E$.  
Eq.~\eqref{eq:kpole} then leads to 
\begin{align}
2m_\star \epsilon =\left(\frac{1}{a}+{m_\star \epsilon r_e}\right)^2\,.
\label{eq:Binding}
\end{align}
This formula is remarkable. Take as an example the case of proton-neutron system in the spin-1 configuration. The values quoted above ($a = \unit[5.42]{fm}$ and $r_e=\unit[1.75]{fm}$), which characterize the velocity dependence of the cross section $\sigma_{pn}$, can be used to solve for  the binding energy of the deuteron.  The result is in perfect agreement with the observed value of $\epsilon=\unit[2.2]{MeV}$.

Nonetheless, not every pole is related to a bound state. The latter are only associated with $k^\text{pole} = i|k|$ ({\it i.e.}\/, ${\rm Im}\ k > 0$).  
Poles with a negative imaginary part correspond to either virtual levels ($k^\text{pole} = - i|k|$) or resonances ($k^\text{pole} = \kappa_d - i|\gamma_d|$).\footnote{\label{fn:poles}For $\kappa_d \neq 0$, the imaginary part of $k^\text{pole}$ cannot be positive to conserve the total probability, see e.g.~\cite{Sitenko:102667}. } An example of the former is given by the collision of  neutrons ($a = \unit[-18.9]{fm}$ and $r_e=\unit[2.75]{fm}$). No bound state of two neutrons exists in nature. In fact,  the state inducing such scattering is  a virtual level.

The relevance of this for SIDM is that if DM forms a bound state, as predicted in many well-motivated scenarios, the corresponding binding energy would be related to the parameters that determine velocity dependence of the self-interaction cross section by means of Eq.~\eqref{eq:Binding}. 
This is particularly true for the Yukawa potential. At the peaks of the cross section (see e.g. Fig.~\ref{fig:alpha}), we found that $|a|\to \infty$. Eqs.~(\ref{eq:kpole}-\ref{eq:Binding})  in turn suggest $|a| \sim 1/\sqrt{2m\epsilon}$ with $\epsilon\to 0$. The peaks in the cross section are thus related to the existence of nearly zero-energy  bound states. 

Even though the regime associated with such bound states is usually referred to as ``resonant regime'',  we would like to emphasize that there are no intermediate particles produced on shell, {\it i.e.}\/, particle resonances. Instead, there are parametric resonances in the sense that for certain parameter combinations the cross section saturates the unitarity limit, where $\epsilon$ approaches zero.  All this explains why the presence of poles with very small $\epsilon$ affects the scattering cross section dramatically.  
\subsection{Resonant SIDM}

In the case of a particle resonance mediating the self-scattering, it is straightforward to see that the kinetic energy $E=(k^\text{pole})^2/(2m_\star)$ is complex, with its real and imaginary parts corresponding to the energy above the   threshold $E_R$, and the decay width of the resonant state $\Gamma(E)$, respectively. More precisely, for the $\ell=0$ case,  $E = E_R- i\Gamma(E)/2$, which  together with Eq.~\eqref{eq:EResonant} leads to the well-known formulas
\begin{eqnarray}
\delta_0 &=& \tan^{-1}\left({ \Gamma(E)/2 \over E_R -E} \right)\,,  \\
\sigma _0 &=& \frac{4\pi }{m E}\frac{\Gamma(E)^2/4}{\left(E-E_R\right)^2+\Gamma^2(E)/4}\,.
\end{eqnarray}

\vspace{.2cm}

To conclude, when the scattering is induced by a bound state, a virtual level or a resonance, this shows up as momentum poles in the complex $k$ plane. Depending on the sign of  the scattering length and the ratio $r_e/a$,  the effective range theory allows to predict which one actually takes place. In fact, one can elaborate further on  the nature of the intermediate state in the scattering process using $r_e$ and $a$. For instance, in the context of the deuteron, Weinberg showed  that one can infer whether the intermediate state is composite or not from the sign of the effective range~\cite{Weinberg:1965zz}.  
Discussing these interesting topics lies beyond the scope of this work.

\subsection{SIMPs} 

The Strongly Interacting Massive Particle (SIMP) is a proposal where the thermal freeze-out occurs by a $3\rightarrow 2$ transition, which is important when the dynamics is strongly coupled, hence the name \cite{Hochberg:2014dra}.  It can be naturally realized in QCD-like gauge theories where pions interact via the Wess--Zumino--Witten term \cite{Hochberg:2014kqa}.  Many variations and mediation mechanisms are discussed in the literature \cite{Yamanaka:2014pva,Lee:2015gsa,Hansen:2015yaa, Bernal:2015xba, Bernal:2015bla, Bernal:2015ova, Choi:2015bya, Hochberg:2015vrg, Kuflik:2015isi,   Choi:2016hid, Choi:2016tkj, Pappadopulo:2016pkp,Farina:2016llk, Dey:2016qgf,Cline:2017tka, Choi:2017mkk, Choi:2017zww,Chu:2017msm, Choi:2018iit, Berlin:2018tvf, Choi:2019zeb, Bhattacharya:2019mmy}.

The SIMP mechanism prefers dark matter mass in the range from 100~MeV to GeV, and is in marginal conflict with the cluster data as seen in Fig.~\ref{fig:contours}.  On the other hand, the strong dynamics often leads to existence of real resonances, bound states, and/or virtual levels, which can improve the agreement by suppressing the cross section at high velocities.  In fact, such resonances in QCD-like models of SIMPs are possible \cite{MMT}.

\section{Improving the effective-range approximation}
\label{sec:gen}

Although Figs.~\ref{fig:approxRepul} and \ref{fig:approx} show that the effective-range approximation works remarkably well in large portions of the parameter space of the Yukawa potential, they make clear that  the approximation fails close to the parameter points where the cross section vanishes, {\it i.e.}\/, at the anti-resonances. In fact, for realistic $S$-wave Breit-Wigner resonances, it may not work for all possible values of the momentum. Likewise, so far we have used the effective range approach to discuss the non-relativistic DM scattering  induced by short-range interactions while inelastic scatterings have been neglected  
to make sure that the potential, as well as the phase shift, is always real. 
In this section, we demonstrate that all these effects can be properly described by extending the effective range formalism.

 \begin{figure}[t]
\includegraphics[trim=0.cm 0cm 0.0cm 0cm,clip,width=\textwidth]{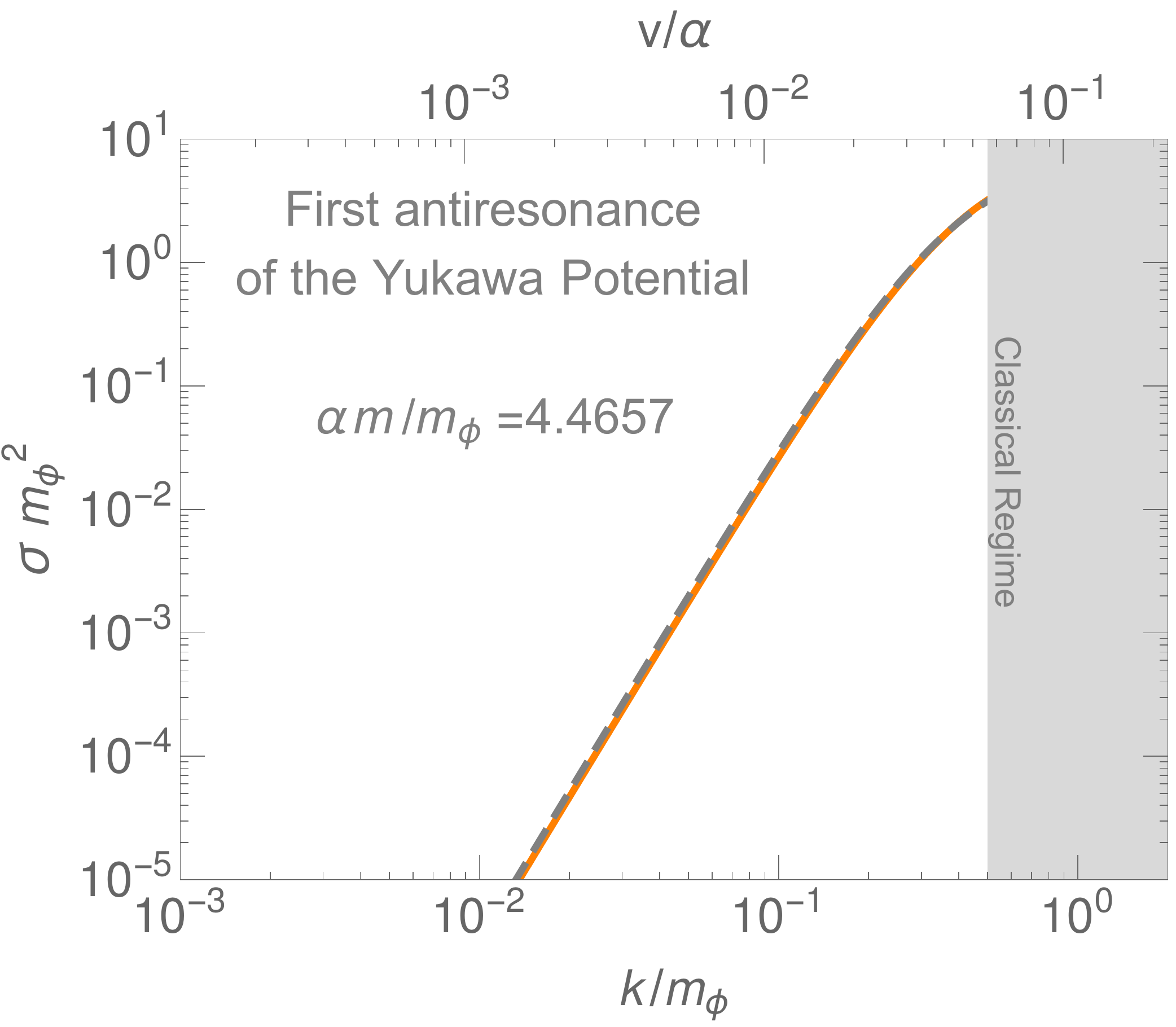}
\caption{ $S-$wave cross section for the first antiresonance of the attractive Yukawa potential (solid-orange) and the approximation (dotted-gray) based on the improved effective-range formula (Eq.~\eqref{eq:improved_ert}) giving $\mathfrak{a} \approx R =-0.85$ and $\mathfrak{r_e}=13.4$. For comparison, the standard effective-range approximation gives $a=0$  and therefore  a negligible cross section everywhere, which does not show up in the plot.   }
\label{fig:threepara}
\end{figure}

\subsection{Antiresonances}

$S$-wave antiresonances  are probably the simplest example where the effective range formalism fails. In contrast to the prediction of Eq.~\eqref{eq:swave_case}, the scattering amplitude and the cross section vanish at a particular value of the momentum, but not everywhere. One possible way to account for this is to decompose the total phase shift into two pieces --one of them satisfying the effective range approximation-- in such a way that they interfere destructively.  More precisely, 
\begin{equation}
	e^{2i\delta_0(k)} = e^{2i k R }  e^{2i \delta_a(k) }= e^{2i k R }\left( \,{ -{1\over \mathfrak{a} }+ {1\over 2} \mathfrak{r_e} k^2 +ik \over -{1\over \mathfrak{a} }+ {1\over 2} \mathfrak{r_e} k^2-ik } \right)\,,
\label{eq:improved_ert}
\end{equation}
which leads to the scattering amplitude  
 \begin{equation}\label{eq:3para}
	f_0(k) =  {e^{2i\delta_0(k)} -1 \over 2 i k}=   {e^{2ikR} -1 \over 2 i k}+    \,{e^{2ikR}  \over -{1\over \mathfrak{a} }+ {1\over 2} \mathfrak{r_e} k^2-ik }\,,
\end{equation}
vanishing  at certain values of the momentum, as required.    Note that here ${\mathfrak{a} }$ and $\mathfrak{r_e}$ are not the standard scattering length and effective range. 

At small $k$, this is equivalent to   $f_0(k) = R+ (-1/\mathfrak{a} +\mathfrak{r_e} k^2/2-i k )^{-1}$, as suggested in $\S$~134 of Ref.~\cite{Landau:1991wop}.   
However,   the latter expression does not respect unitarity because $|e^{2i\delta_0(k)}| \neq 1$ as follows from Eq.~\eqref{eq:fkdef}. In contrast, our parametrization of Eq.~\eqref{eq:improved_ert} respects unitarity manifestly. 
 
Fig.~\ref{fig:threepara} illustrates our parametrization for the first antiresonance of the attractive Yukawa potential. The difference between the numerical result and the approximation based on Eq.~\eqref{eq:improved_ert} is imperceptible.

\subsection{Sharp resonances}

\begin{figure}[t]
\includegraphics[trim=0.cm 0cm 0.0cm 0cm,clip,width=\textwidth]{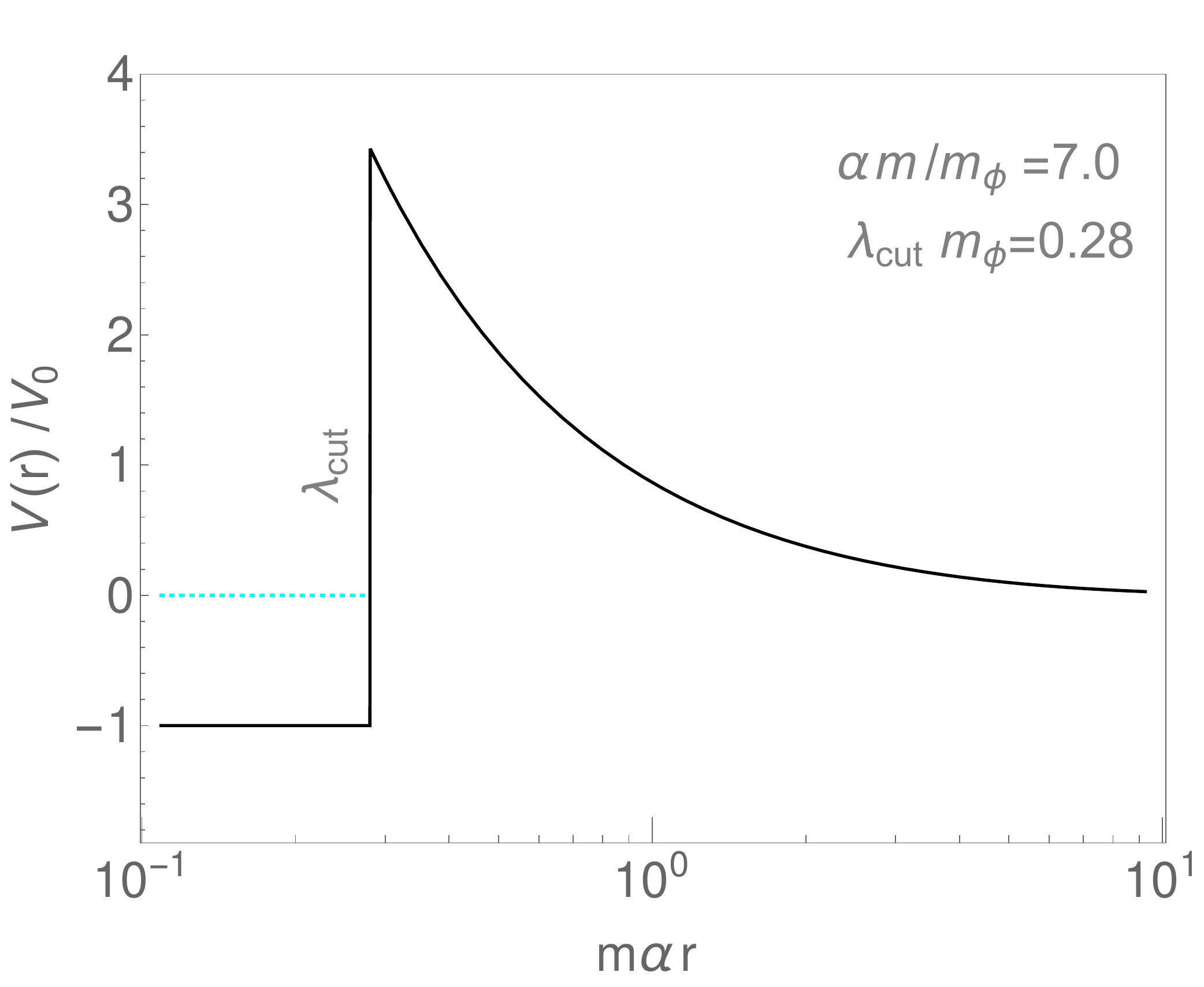}\\
\caption{Potential in Eq.~\eqref{eq:pot_well}, which exhibits unstable bound-states, that is, real resonances.}
\label{fig:threepotential}
\end{figure}

\begin{figure*}[t]
\includegraphics[trim=0.cm 0cm 0.0cm 0cm,clip,width=0.48\textwidth]{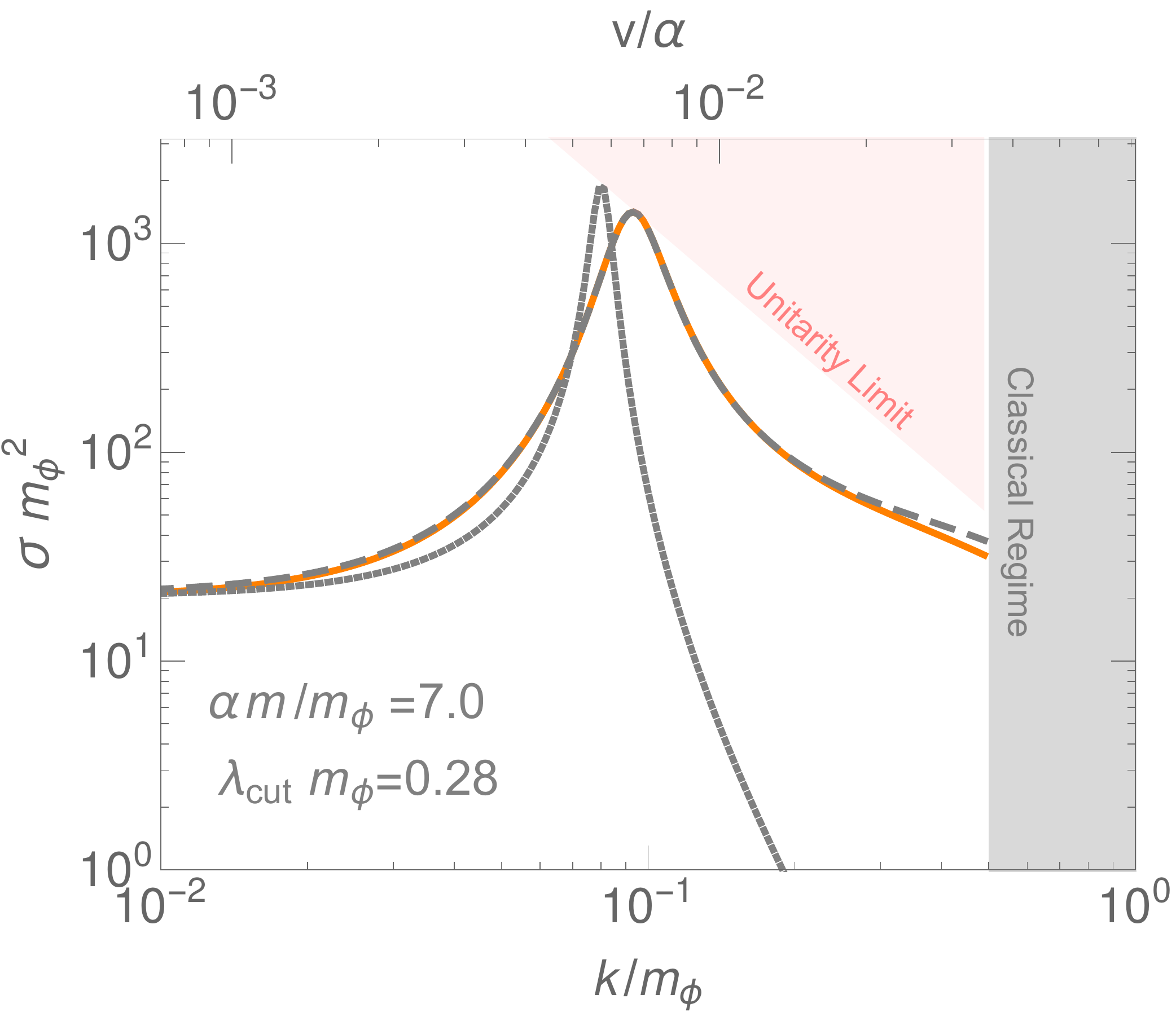}
\includegraphics[trim=0.cm 0cm 0.0cm 0cm,clip,width=0.48\textwidth]{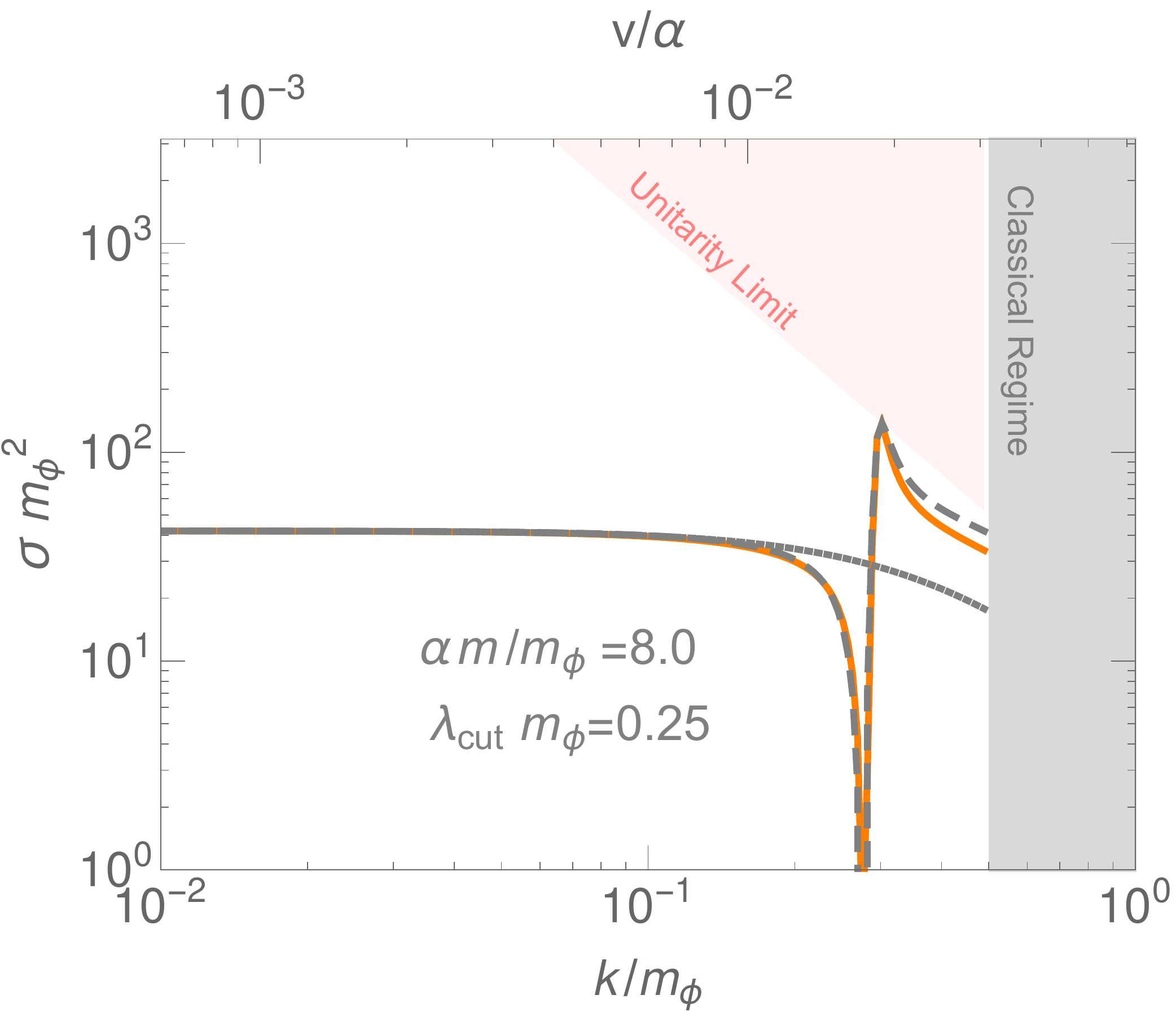}
\caption{Same as Fig.~\ref{fig:threepara} but for the potential of Eq.~\eqref{eq:pot_well}.  The exact result is the solid orange line, the dashed line is the approximation based on Eq.~\eqref{eq:improved_ert} while the dotted gray line is the standard effective-range approximation of Eq.~\eqref{eq:swave_case}. }
\label{fig:threepara2}
\end{figure*}

 Eq.~\eqref{eq:improved_ert}  can also describe sharp resonances (second term) accompanied with a continuum piece (first term),  while the standard effective range approximation can only describe one of the two. 
To illustrate this fact, let us consider the potential
\begin{equation}
V(r) = - m\alpha^2 \Theta(\lambda_\text{cut}-r) + \Theta(r-\lambda_\text{cut}) \frac{\alpha}{r}e^{-m_\phi r},
\label{eq:pot_well}
\end{equation}
which we depict in Fig.~\ref{fig:threepotential} for a particular parameter choice.

This potential gives rise to positive-energy bound states that decay through quantum tunneling. These are real resonances in contrast to the peaks associated with the Yukawa potential in Fig.~\ref{fig:alpha}, as explained above. In fact, as shown in the left panel of Fig.~\ref{fig:threepara2}, the scattering cross section exhibits a resonant enhancement for certain values of the momentum. These correspond to the formation of unstable bound-states. 
Interestingly, while the standard effective range formula describes the continuum part of cross section fairly well, it fails to describe scattering cross section at the peak. In spite of this, the improved formula approximates the exact result very well. Notice that, as explained above, such an improvement is in practice adding a continuum piece to cross section (except for negative and positive interference just below and above the resonance), which has already been considered in phenomenological studies of resonant SIDM~\cite{Chu:2018fzy}.  

In the same fashion, Eq.~\eqref{eq:improved_ert} can simultaneously describe a resonance and an antiresonace as it is the case for potential in Eq.~\eqref{eq:pot_well} for certain points of the parameter space. This is  shown in the right panel of Fig.~\ref{fig:threepara2} for another parameter choice. In this case, the antiresonance is  induced by the destructive inferences between the continuum and resonance parts. %

\subsection{Inelastic scatterings } 
\label{subsec:inelastic}

Even though we have assumed that inelastic scatterings --such as DM annihilation or radiative capture~\cite{MarchRussell:2008tu,Braaten:2013tza}-- are subleading, in principle they can play a role. For instance, in the vanilla light-mediator model, DM typically annihilates into the mediators, leading to a complex potential~\cite{Blum:2016nrz}. 
In this case the corresponding phase shift can be decomposed as $\delta_\ell = \text{Re}\, \delta_\ell+ i  \text{Im} \,\delta_\ell$. Assuming $|\delta_\ell| \gg  \text{Im} \delta_\ell \ge 0$, one can adopt the $S$-wave effective range approximation in the following way
\begin{equation}
	k\cot\delta_0 \simeq k \cot(\text{Re} \delta_0) - {i  k  \text{Im} \delta_0 \over \sin^2(\text{Re} \delta_0)}  \simeq  - {1\over a } + {r_e \over 2}k^2\,,
\end{equation}
where $a$ and $r_e$ contain a subleading imaginary component~\cite{PhysRevA.66.022716}. If we further neglect the imaginary component of $r_e$, the $S$-wave annihilation and scattering cross sections are related via  
\begin{equation}
	\sigma_{an,\,0}(k) = {4\pi \over k^2} {1 - |e^{2i\delta_0}|^2\over 4 }\simeq {\sigma_0(k) \over k }  {|\text{Im}  a| \over (\text{Re}a)^2 }  \,.
\end{equation}
Note that such expression does not violate the unitarity limit as its last factor  vanishes at $|a|\to \infty$. Moreover, it shows that  $\sigma_{an,\,0}(k)\,k$ becomes constant at  $k\ll 1/|a|$ as long as the effective range approximation applies. 
 For more discussions on parametrizing the relation between elastic and inelastic cross sections, see e.g. \cite{Braaten:2013tza, Blum:2016nrz}.

\section{Summary and Outlook}%
\label{sec:conclusions}

In this work, we have studied the effective range approach as a model-independent way to parametrize DM scattering cross sections in astrophysical halos. While it only contains two parameters besides the DM mass, it provides a good description of the self-scattering in most  appealing SIDM  scenarios, including SIMPs, SIDM with a light mediator and resonant SIDM models.

Starting with a brief introduction to the effective-range approach, we have studied the astrophysical implications. 
In general, there exists a velocity scale, $(m|a|)^{-1}$, below which the scattering cross section can be treated as a constant. For velocities well above this scale, the cross section quickly decreases. 
Taking bounds derived from current cluster observations, we have reached the conclusion that DM masses below several GeV are excluded for $\sigma/m\sim 10$~cm\,$^2$/g in dwarf-sized halos. See Figs.~\ref{fig:con} and ~\ref{fig:contours}. The tentative non-vanishing values of $\sigma/m$ --extracted from observational data at various scales-- can also be fit in terms of the effective-range parameters. See Fig.~\ref{fig:fit}. Our results suggest that   more precise measurements and better extractions are needed to identify or constrain such parameters.  Nevertheless, since the value of $r_e/a$ only affects the cross section around $v\sim 1/(m|a|)$, obtaining its value from future observations might be challenging.

In addition, we  have further investigated the correspondence of the scattering length $a$ and the effective range $r_e$ to the model parameters of several popular SIDM scenarios. In general, the scattering cross sections calculated from $a$ and $r_e$  agrees well with the exact values as long as the range of the interaction is sufficiently short. Moreover, the effective-range approach  demonstrates that significant enhancements in the self-scattering cross section are induced by the poles in the complex plane of the DM momentum. In analogy to nuclear physics, such poles can be interpreted as intermediate physical states, such as a bound state, a virtual level or a resonance.

In the end, we have briefly commented on possible extensions of the  effective-range approach, especially for the cases that contain anti-resonances or sharp resonances. Besides,  we have also shown that  
it is possible to  study subleading inelastic processes such as DM annihilations using the same framework.  

We believe the effective-range approach provides a simple, yet very useful,  parametrization to consistently take into account the velocity dependence of DM self-interactions in cosmological simulations involving different astrophysical scales. For instance, such velocity-dependence may play an important role in better understanding the  evolution of the sub-halos that move inside the Milky Way halo.  This is left  for future work.

\vspace{.1cm}

{\renewcommand{\addtocontents}[2]{}
\section*{Acknowledgements}
}

We thank  Kai Schmidt-Hoberg for   discussions.  X.C. is supported by the `New Frontiers' program of the Austrian Academy of Sciences. C.G.C. is supported by the ERC Starting Grant NewAve (638528). X.C. and C.G.C. thank the Erwin Schr\"odinger International Institute for  hospitality while this work was completed.   
 H.M. thanks the Alexander von Humboldt Foundation for support while this work was completed.  H.M. was supported by the NSF grant PHY-1638509, by the U.S. DOE Contract DE-AC02-05CH11231, by the JSPS Grant-in-Aid for Scientific Research (C) (17K05409), MEXT Grant-in-Aid for Scientific Research on Innovative Areas (15H05887, 15K21733), by WPI, MEXT, Japan,   by the Binational Science Foundation  (grant No. 2016153), and by Hamamatsu Photonics K.K.

\vspace{.4cm}

\appendix

\section{The effective range theory}
\label{app:ERT}

The phase shifts associated with the self-scattering of DM particles are obtained by solving  the Schr\"{o}dinger equation for the radial wavefunction $R_{\ell,k}(r)$ of the reduced DM two-particle system. This is given by
\begin{equation}
\frac{1}{r^2} \frac{d}{dr} \Big( r^2 \frac{d R_{\ell,k}}{dr} \Big) + \Big( k^2 - \frac{\ell (\ell + 1)}{r^2} - m V(r) \Big) R_{\ell,k} = 0\,, \label{eq:radial}
\end{equation}
together with a boundary condition demanding that $r R_{l,k}$ must vanish at $r=0$. In fact,  close to the origin it is expected that the angular-momentum term dominates for sufficiently well-behaved potentials, in which case $R_{l,k} \propto r^{l}$. 
At large distances from the origin, the potential vanishes and the wave function must be that of a free particle, {\it i.e.}\/, a superposition of two spherical waves.  The phase shift, $\delta_\ell$, parametrizes such a superposition. More precisely, at $r\to \infty$ the asymptotic behavior $R_{\ell,k}(r)$ is given by
\begin{equation}
\label{eq:Rasymp}
R_{\ell,k}(r) \propto  \cos\delta_\ell \, j_\ell(kr) -  \, \sin \delta_\ell n_\ell(kr)\approx \frac{1}{r} \sin\left(kr-\frac{l\pi}{2}+\delta_\ell\right)  \, ,
\end{equation}
where $j_\ell$ and $n_\ell$ are respectively the spherical Bessel functions of first and second order.

\subsection{A simple method to find the phase shift} 
 \label{phaseshift:solution}

In the SIDM context, Ref.~\cite{Tulin:2013teo} presented a systematic method for solving Eq.~\eqref{eq:radial}. Here we would like to point out a simpler possibility that will not only provide a powerful method to solve for the phase shift but  will also allow us to define the scattering length and the effective range. Let us first define
\begin{equation}
t_{\ell,k} (r) = \frac{j_\ell(k r) \left(\frac{R_{\ell,k}'(r)}{ R_{\ell,k}(r)}-\frac{\ell}{ r}\right)+k\,j_{\ell+1}(k r)}{n_\ell(k r) \left(\frac{R_{\ell,k}'(r)}{ R_{\ell,k}(r)}-\frac{\ell}{ r}\right)+k\, n_{\ell+1}(k r)}\,.
\end{equation}
Simple algebra shows that
\begin{equation}
\frac{d t_{\ell,k} (r)}{dr} = -k\, m\, r^2 V(r) \left(j_\ell(k r)-t_{\ell,k}(r) n_\ell
(k r)\right)^2\,.
\label{eq:radial1st}
\end{equation}
The fact that $R_{\ell,k} \propto r^{\ell}$ and Eq.~\eqref{eq:Rasymp} fix the boundary conditions of this differential equation to
\begin{align}
t_{\ell,k} (0) = 0 &&\text{and}&& t_{\ell,k}(r)\to \tan \delta_\ell\quad \text{at}\quad r\to\infty\,.
\end{align} 

Notice that $j_\ell (kr)\propto k^{\ell} $ and $n_\ell (kr)\propto k^{-(\ell+1)} $ in the limit $k\to 0$, which  together with Eq.~\eqref{eq:radial1st} imply that $\tan \delta_\ell \propto k^{2\ell+1}$ for small momenta. The corresponding coefficient of proportionality defines scattering length $a_{\ell}$. More precisely, 
\begin{equation}
a_{\ell}^{2\ell+1} \equiv -\lim_{k \to 0} \frac{\tan\delta_\ell}{k^{2\ell+1}} \,.
\end{equation}
The function $k^{2\ell+1} \cot \delta_\ell$ is thus analytic at $k=0$. The next-to-leading term determines the effective range, $r_{e,\ell}$, by means of
\begin{equation}
k^{2\ell+1} \cot \delta_\ell=  -\frac{1}{a_{\ell}^{2\ell+1}} + \frac{1}{2 \,r_{e,\ell}^{2\ell-1}} k^2+{\cal O}(k^4)\,.
\end{equation}

As a by-product we have found a powerful method to solve for the phase shift.\footnote{Simple changes of variable on Eq.~\eqref{eq:radial1st} allow to simplify the method further. For instance, in the presence of resonances --for which the angle $\delta_\ell$ goes beyond $\pi/2$ and its tangent takes values in different branches-- it is more convenient to use 
\begin{equation}
\frac{d \delta_{\ell,k} (r)}{dr} = -k\, m\, r^2 V(r) \left(\cos \delta_{\ell,k}(r) j_\ell(k r)-\sin \delta_{\ell,k}(r) n_\ell
(k r)\right)^2\,,
\label{eq:radial2nd}
\end{equation}
with
\begin{align}
\delta_{\ell,k} (0) = 0 &&\text{and}&& \delta_{\ell,k}(r)\to \delta_\ell\quad \text{at}\quad r\to\infty\,.
\label{eq:BC2}
\end{align} 
Alternatively, using the spherical Hankel function of first kind, $h^{(1)}_{\ell}$, one finds the even simpler formula
\begin{equation}
\frac{d \delta_{\ell,k} (r)}{dr} = -k\, m\, r^2 V(r) {\text{ Re}}\left[ e^{i \delta_{\ell,k}(r) }h^{(1)}_{\ell}(kr)\right]^2\,,
\label{eq:radial3nd}
\end{equation}
which must be solved together with Eq.~\eqref{eq:BC2}. Due to the fact that $h^{(1)}_{\ell} (r) \propto r^{-(\ell+1)}$, this algorithm is numerically unstable for very large $\ell$. Since those values are only relevant in the classical regime, in that case it might be more efficient to use the analytical formula for $\delta_\ell$ from classical physics~\cite{Landau:1991wop}. 
}
In fact, it is numerically much more efficient to integrate Eq.~\eqref{eq:radial1st} than to integrate Eq.~\eqref{eq:radial}, not only because the former is of first order but also because solving Eq.~\eqref{eq:radial1st} does not require matching the solution to a plane wave at infinity in order to find the phase shift.

Eq.~\eqref{eq:radial1st}  can be solved by expanding on the potential, with the first term determining the Born regime, which is given by
\begin{equation}
\label{eq:Born}
\tan \delta_l\Bigg|_\text{Born} = -k\,m \int^\infty_0 r^2 V(r) j_\ell(kr)^2 dr\,.
\end{equation}

\subsection{The $S$-wave case} 

Let us take $\ell=0$ and introduce $u_k(r) =rR_{k\,,0}(r)$. Then, Eqs.~\eqref{eq:radial} and \eqref{eq:Rasymp} read
\begin{align}
\left(\frac{d^2}{dr^2}+ k^2- m V(r) \right)u_k(r) =0\,,
\label{eq:u}
\end{align}
and
\begin{align}
u_k(0)=0\,,~ u_k(r)\to \psi_k(r) =  \frac{\sin \left(kr +\delta_0\right)}{\sin \delta_0} \text{~at~} r\to\infty\,.
\label{eq:u}
\end{align}
Here we have chosen a convenient  normalization factor for $u_k$.  In the following we will find it useful to employ the previous definition of $\psi_k(r)$  for any positive value of $r$. Simple algebra proves that for any potential
\begin{equation}
u_k (r)\frac{du_0(r)}{dr}- u_0(r)\frac{du_k(r)}{dr} \Bigg|^r_0= k^2 \int^r_0 u_0(r') u_k(r') dr'\,.
\label{eq:rel1}
\end{equation} 
Moreover, using the fact that $\psi_k (r)$ is the solution of the Schr\"odinger equation for $V(r)=0$, we find that
\begin{equation}
\psi_k (r')\frac{d\psi_0(r')}{dr}- \psi_0(r')\frac{d\psi_k(r')}{dr}\Bigg|^r_0 = k^2 \int^r_0 \psi_0(r') \psi_k(r') dr'\,.
\label{eq:rel2}
\end{equation} 
Notice that $\psi_0(r) =1-r/a_0$, where $a_0$ is the scattering length. Subtracting Eq.~\eqref{eq:rel1} from  Eq.~\eqref{eq:rel2}, taking  $r\to\infty$ and using the fact that $u_k$ and $\psi_k$ approach to each other in that limit, we find that
\begin{eqnarray}
k\cot \delta_0&=&-\frac{1}{a_0}+k^2 \int^\infty_0 \left( \psi_0\psi_k- u_0 u_k\right)dr \notag\\
&=&-\frac{1}{a_0} +\frac{1}{2}r_{e,0} k^2 +{\cal O}(k^4) \,.
\label{eq:Bethe}
\end{eqnarray}
where
\begin{equation}
r_{e,0} = 2 \int^\infty_0 \left( \psi_0^2- u_0^2\right)dr\,.
\label{eq:reBethe}
\end{equation}
This is the original expression found by Bethe~\cite{Bethe:1949yr}, who argued that the expansion in Eq.~\eqref{eq:Bethe} approximates the phase shift with a great accuracy because $\psi_k$ and $u_k$ differ only  where the potential is non-negligible. Note that this conclusion is based on the assumption that in this region both wave functions depend very weakly on  $k$, which is generally true, when the potential energy is much larger than kinetic energy and $kr$ is small.  
  
To qualitatively understand the effective range, one can consider the following upper bound, which is valid  for potentials that effectively vanish at distances greater than certain range $R$~\cite{Phillips:1996ae}, as it is the case of  the Yukawa potential. Then, $\psi_0$ and $u_0$ in Eq.~\eqref{eq:reBethe} coincide for $r\gtrsim R$, which  implies that
\begin{align}
  	{r_{e,0} \over R} & \approx  \frac{2}{R} \int^R_0 \left( \left(1-\frac{r}{a_0}\right)^2- u_0^2\right)dr \notag\\ & \leq   2 \left( 1- { R \over a_0} + {1\over 3}\left({R\over a_0 }\right)^2 \right)\,.
  \end{align}
Moreover, for shallow attractive potentials, $u_0$ behaves like a slowly-varying sine function, where mostly $u_0\lesssim \psi_0$, resulting in a positive $r_{e,0}$. See also $\S$.~133 of \cite{Landau:1991wop}.

\subsection{The Hulth\'{e}n Potential}

In the main text, it has been mentioned that the Hulth\'{e}n potential
\begin{equation}
	V(r)=\pm \alpha \delta e^{-\delta r}/(1-e^{-\delta r})
\end{equation}
approximates well the Yukawa potential if one sets $\delta=\sqrt{2\zeta(3)} m_\phi$, where $\alpha$ gives the coupling and $m_\phi$ is the mediator mass of the Yukawa potential~\cite{Cassel:2009wt}. 
The advantage of employing  the Hulth\'{e}n potential is that its corresponding Schr\"{o}dinger equation is analytically solvable, and yield the $S$-wave phase shift~\cite{Cassel:2009wt, Tulin:2013teo}
\begin{equation}
	\delta_0 = \arg \left(  {i\Gamma (\lambda_+ + \lambda_- -2 ) \over \Gamma (\lambda_+)\Gamma (\lambda_-)} \right)\,.
\end{equation}
Here, the dimensionless function $\lambda_\pm $ is given by  $1+ i m v/(2\delta) \pm \sqrt{ \alpha m/\delta -  m^2 v^2 /(2\delta)^2}$.

\begin{figure}[t]
\centering
\includegraphics[trim=0cm 0cm 0.0cm 1cm,clip,width=.97\textwidth]{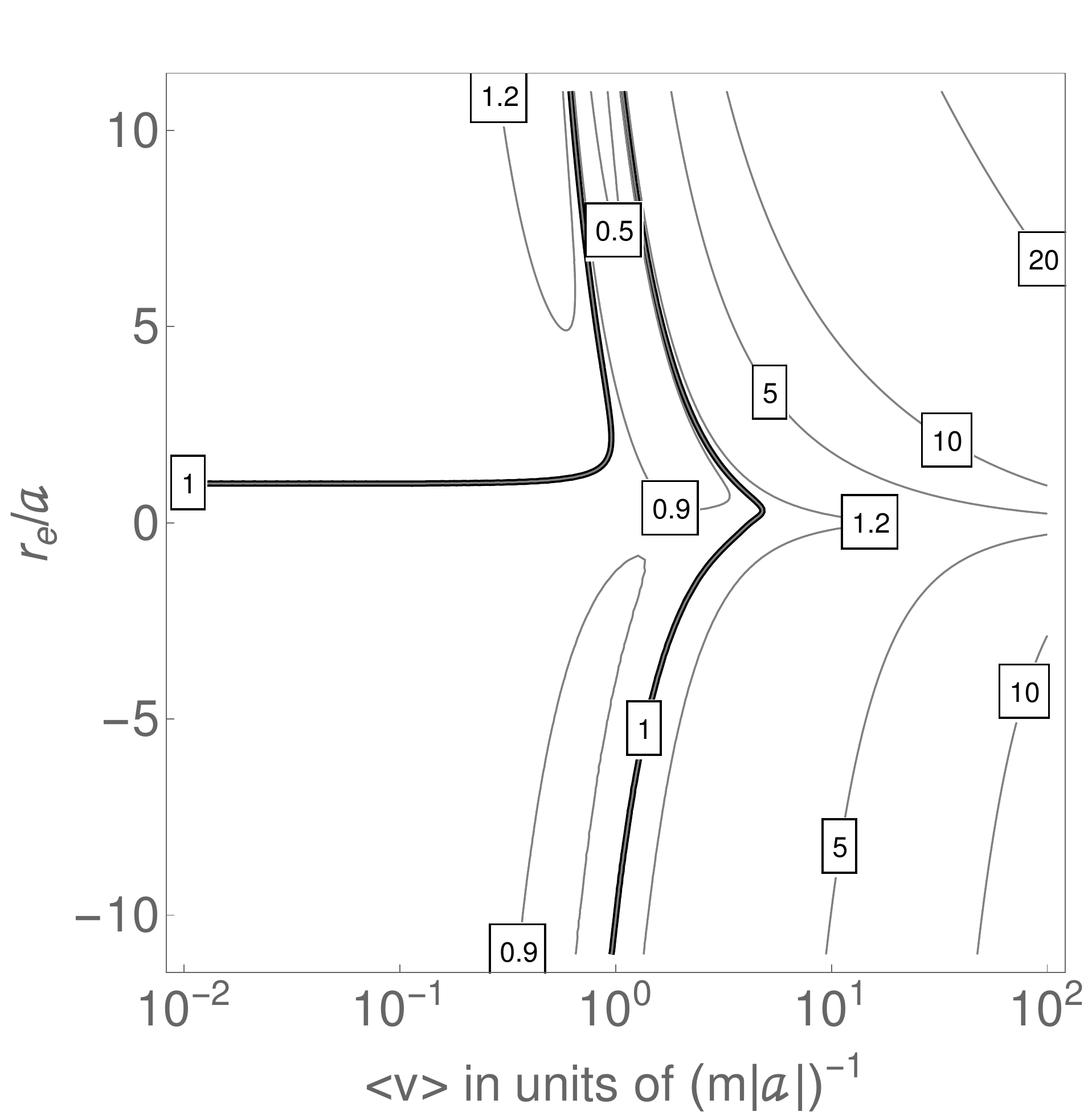}
\caption{Contours of the ratio of $\langle \sigma v \rangle $ and $\sigma (\langle v\rangle) \langle v\rangle $ as a function of $a m \langle v\rangle$ and $r_e/a$. 
}
\label{fig:wavefunction}
\end{figure}

Using Eqs.~\eqref{eq:Bethe} and \eqref{eq:reBethe}, one can obtain the analytical expressions of the 
$S$-wave effective-range parameters as
\begin{eqnarray}
	\hspace{-100pt}a   &=& {\psi^{(0)}(1+\eta) + \psi^{(0)} (1-\eta ) +2\gamma \over \delta}\,,\\
	\hspace{-100pt}r_e  &=& {2a \over 3} -   { 1 \over 3 \delta  \eta\, \left[\psi^{(0)}(1+\eta) + \psi^{(0)} (1-\eta ) +2\gamma \right]^2  }\notag \\
	&& \times \left\{ 3  \left[ \psi^{(1)}(1 +\eta) -  \psi^{(1)}(1 - \eta )  \right] \right. \notag \\ &&+ \left. \eta \left[ \psi^{(2)}(1 + \eta) + \psi^{(2)}(1 - \eta ) +16\zeta(3) \right]  \right\},~~
\end{eqnarray}  
where $\eta=\sqrt{\alpha m/\delta } $,  $\psi^{(n)}(z)$ are the polygamma functions of order $n$ and  $\gamma \simeq 0.5772$ is the  Euler-Mascheroni constant.

\section{Velocity-averaged cross sections}
\label{app:averageV}

The averaged cross section $\langle \sigma v\rangle $ can be calculated in terms of 
\begin{align}
\frac{\langle \sigma v\rangle }{\langle v\rangle}&\equiv\frac{ \int^{\infty}_0 f(v,v_0)\, \sigma v \, dv}{\int^{\infty}_0 f(v,v_0)\, v \, dv} \notag\\ & =\pi a^2    \, \frac{r_e^2\,  (z_+-z_- ) }{ a^2 -  2a r_e } \, \left( \phi(z_+)-\phi(z_-)\right) \,, 
\label{eq:formula}
\end{align}
with
\begin{align}
 \phi (z) &= z\,e^{-z} \Gamma(0,-z)\,,\\  z_\pm &=\frac{32}{\pi( a m \langle v \rangle  )^2} \left( 1-\frac{a}{r_e}\pm\frac{a}{r_e}\sqrt{1- \frac{2r_e}{a}}\right)\frac{a}{r_e}  \,.
\end{align}

We also show the ratio of $\langle \sigma  v\rangle $ and $\sigma (\langle v\rangle) \langle v\rangle$ as a function of $a m \langle v\rangle$ and $r_e/a$ in  Fig.~\ref{fig:wavefunction}, which  shows that within the effective-range approach both coincide at one percent level, except for  the large-velocity regime $\langle v\rangle \gg (|a| m )^{-1}$, where $\sigma$ can be   sensitive to $v$.

{\renewcommand{\addtocontents}[2]{}
\bibliographystyle{utcaps_mod}
\bibliography{ref}
}

\end{document}